\documentclass[final,3p,times,onecolumn,sort&compress,hidelinks]{elsarticle}

\usepackage[normalem]{ulem}
\usepackage[utf8]{inputenc}
\usepackage{slashed}
\usepackage{graphicx}
\usepackage{hyperref}
\usepackage{bm}
\usepackage[usenames]{color}
\usepackage{listings}
\usepackage{pdfpages}
\usepackage{pstricks}
\usepackage{color}
\usepackage{enumerate}
\usepackage{float}
\usepackage{amsmath}
\usepackage{amssymb}
\usepackage{mathtools}
\usepackage{amssymb}
\usepackage{marvosym}

\newcommand{\nctmo}{\frac{1}{N_c^2-1}}
\newcommand{\ths}{\hat{s}}
\newcommand{\tht}{\hat{t}}
\newcommand{\thu}{\hat{u}}

\usepackage{soul}

\usepackage{amsmath,amssymb}
\usepackage{bm,bbm}
\usepackage{graphicx, graphics, color}
\usepackage{slashed}
\usepackage{booktabs}
\allowdisplaybreaks

\newcommand{\bea}{\begin{eqnarray}}
\newcommand{\eea}{\end{eqnarray}}

\newcommand{\ben}{\begin{eqnarray}}
\newcommand{\ee}{\end{equation}}
\newcommand{\een}{\end{eqnarray}}

\newcommand*{\FigPath}{./figs/}%

\begin{document}

\begin{frontmatter}

\author[label1]{Leonard Gamberg}\ead{lpg10@psu.edu}
\author[label2,label3,label4]{Zhong-Bo Kang}\ead{zkang@physics.ucla.edu}
\author[label1]{Daniel Pitonyak}\ead{dap67@psu.edu}
\author[label1,label5]{Alexei Prokudin} \ead{prokudin@jlab.org}
\address[label1]{Division of Science, Penn State University Berks, Reading, Pennsylvania 19610, USA}
\address[label2]{Department of Physics and Astronomy, University of California, Los Angeles, California 90095, USA}
\address[label3]{Mani L. Bhaumik Institute for Theoretical Physics, University of California, Los Angeles, California 90095, USA}
\address[label4]{Theoretical Division, Los Alamos National Laboratory, Los Alamos, New Mexico 87545, USA}
\address[label5]{Theory Center, Jefferson Lab, 12000 Jefferson Avenue, Newport News, Virginia 23606, USA}

\title{Phenomenological constraints on $A_N$ in \\ $p^\uparrow p\to \pi\, X$ from Lorentz invariance relations}

\begin{abstract}
We present a new analysis of $A_N$ in $p^\uparrow p\to \pi\, X$ within the collinear twist-3 factorization formalism.  We incorporate recently derived Lorentz invariance relations into our calculation and focus on input from the kinematical twist-3 functions, which are weighted integrals of transverse momentum dependent (TMD) functions. In particular, we use the latest extractions of the Sivers and Collins functions with TMD evolution to compute certain terms in $A_N$. Consequently, we are able to constrain the remaining contributions from the lesser known dynamical twist-3 correlators.   
\end{abstract}

\begin{keyword}
transverse spin \sep perturbative QCD \sep collinear factorization \sep JLAB-THY-17-2405 

\PACS 12.38.-t \sep 12.38.Bx \sep 13.75.Cs \sep 13.85.Ni \sep 13.88.+e  


\end{keyword}

\end{frontmatter}

\date{\today}

\section{Introduction \label{sec:intro}}

The endeavor to probe the spin structure of the proton through transverse single-spin asymmetries (TSSAs), denoted $A_N$, in high-energy single inclusive lepton-hadron and hadron-hadron scattering processes, i.e., $A^\uparrow+B\to C+X$~\footnote{One could also have C transversely polarized instead of A.}, has received considerable attention from both the experimental and theoretical communities~\cite{Pitonyak:2016hqh,Aschenauer:2016our}. For the case where the produced particle $C$'s transverse momentum $P_{CT}\gg \Lambda_{QCD}$, TSSAs manifest themselves as sub-leading twist (twist-3) effects calculable within perturbative QCD (pQCD). The computational techniques and methodology of this collinear twist-3 factorization framework were developed rigorously in Refs.~\cite{Efremov:1981sh,Efremov:1984ip,Qiu:1991pp,Qiu:1991wg,Qiu:1998ia,Kanazawa:2000hz,Kouvaris:2006zy,Eguchi:2006qz,Eguchi:2006mc,Koike:2006qv,Koike:2007rq,Zhou:2008fb,Koike:2009ge,Yuan:2009dw,Kang:2010zzb,Beppu:2010qn,Metz:2010xs,Koike:2011mb,Koike:2011nx,Metz:2012ct,Kanazawa:2013uia,
Beppu:2013uda,Kanazawa:2014nea,Kanazawa:2015ajw,Koike:2015zya}.  Over the last 40 years, there have been many measurements of large TSSAs~\cite{Klem:1976ui,Bunce:1976yb,Adams:1991rw,Krueger:1998hz,
Allgower:2002qi,Adams:2003fx,Adler:2005in,Lee:2007zzh,Abelev:2008af,Arsene:2008aa,Adamczyk:2012qj,Adamczyk:2012xd,Bland:2013pkt,Adare:2013ekj,Adare:2014qzo,Airapetian:2013bim,Allada:2013nsw}, whose description, therefore, has become a fundamental test of this pQCD formalism.  

Schematically, one writes the (polarized) differential cross section for $A^\uparrow+B\to C+X$ as
\begin{equation} 
d\sigma(S_{T}) = \,H\otimes f_{a/A(3)}\otimes f_{b/B(2)}\otimes D_{C/c(2)}  + H'\otimes f_{a/A(2)}\otimes f_{b/B(3)}\otimes D_{C/c(2)}  + H''\otimes f_{a/A(2)}\otimes f_{b/B(2)}\otimes D_{C/c(3)}\,,
\label{e:collfac}
\end{equation} 
where $S_T$ is the transverse spin vector of hadron $A$, $f_{a/A(t)}$ is the twist-$t$  parton distribution function (PDF) associated with parton $a$ in hadron $A$ (similarly for $f_{b/B(t)}$), while $D_{C/c(t)}$ is the twist-$t$ fragmentation function (FF) associated with hadron $C$ in parton $c$.  The twist-3 correlators can either be of the 2-parton or 3-parton type and are categorized into intrinsic, kinematical, and dynamical functions~\cite{Pitonyak:2016hqh,Kanazawa:2015ajw}. The intrinsic functions are twist-3 Dirac projections of collinear 2-parton correlators, while the kinematical functions are first transverse momentum moments of transverse momentum dependent (TMD) 2-parton functions. The dynamical functions are 3-parton correlators. The factors $H$, $H'$, and $H''$ are the hard parts for each term, and the symbol $\otimes$ denotes convolutions in the appropriate parton momentum fractions.   One then can calculate the TSSA $A_N$ as
\bea
A_N \equiv \frac{d\Delta\sigma(S_T)}{d\sigma}, 
\,\,\,
{\rm where}\quad
d\Delta\sigma(S_T)\equiv \frac{1}{2}\left[d\sigma(S_T)-d\sigma(-S_T)\right],\,\,\, {\rm and}\quad
d\sigma \equiv\frac{1}{2}\left[d\sigma(S_T)+d\sigma(-S_T)\right].
\label{e:AN}
\eea
In this Letter, we will focus on $A_N$ in the process $p^\uparrow  p\to \pi\,X$,
i.e., single inclusive pion production in collisions between transversely polarized protons and unpolarized protons.

Let us first review the current situation for this observable, which receives contributions from all three terms in Eq.~(\ref{e:collfac}).  That is, the full result involves twist-3 functions from (i) the transversely polarized proton, (ii) the unpolarized proton, and (iii) the (unpolarized) final-state pion. For (i) there are two types of terms that arise:~a  soft-gluon pole (SGP) term and a soft-fermion pole (SFP) term. The former was calculated in Refs.~\cite{Qiu:1998ia,Kouvaris:2006zy} for $qgq$ correlators involving the Qiu-Sterman (QS) function $F_{FT}(x,x)$, as well as in Ref.~\cite{Beppu:2013uda} for tri-gluon ($ggg$) ones.  The latter (SFP) was computed in Ref.~\cite{Koike:2009ge}. We note that $F_{FT}(x,x)$ has an important, model-independent relation~\cite{Boer:2003cm} to the first $k_T$-moment of the TMD Sivers function $f_{1T}^{\perp}(x, \vec{k}_T^{\,2})$~\cite{Sivers:1989cc}. 

The initial attempt to describe $A_N$ in $p^\uparrow  p\to \pi\,X$ did so under the assumption that the QS function was the dominant source of the asymmetry~\cite{Qiu:1998ia,Kouvaris:2006zy}.  However, a series of studies~\cite{Kang:2011hk,Kang:2012xf,Metz:2012ui}
have demonstrated that $F_{FT}(x,x)$ {\it cannot} be the main cause of $A_N$ in this reaction.  Also, the tri-gluon term has been shown to give small effects in the forward region~\cite{Beppu:2013uda} where $A_N$ is most significant. While the SFP piece might play some role, it cannot account for all of the asymmetry~\cite{Kanazawa:2010au,Kanazawa:2011bg}. In addition, term (ii) 
couples two chiral-odd functions:~the quark transversity PDF from the transversely polarized proton and the twist-3 function 
$H_{FU}(x,x)$~\cite{Kanazawa:2000hz,Zhou:2008mz,Kang:2008ey,Kang:2012em} 
from the unpolarized proton, which is related~\cite{Boer:2003cm} to the first $k_T$-moment of the Boer-Mulders function 
$h_1^\perp(x,\vec{k}_T^{\,2})$~\cite{Boer:1997nt}. 
It was shown in~\cite{Kanazawa:2000kp} that such a contribution to $A_N$ is negligible due to the small size of the corresponding hard partonic cross section.

Based on these observations, attention has recently been given 
to term (iii) involving twist-3 effects from partons fragmenting 
into the final-state pion. A few years ago, the complete calculation of this contribution was carried out in Ref.~\cite{Metz:2012ct} and involves three parts:~the kinematical function $H_1^{\perp(1)}(z)$ (i.e., the first $p_\perp$-moment of Collins function $H_1^\perp(z,z^2\vec{p}_\perp^{\,2})$~\cite{Collins:1992kk}\footnote{Studies of the contribution to $A_N$ from the convolution of the transversity and Collins TMD functions in the framework of the so-called Generalized Parton Model are presented in Ref.~\cite{Anselmino:2012rq}.})
the intrinsic function $H(z)$, and the dynamical function \footnote{The dynamical FFs are complex, and we indicate their real and imaginary parts by $\Re$ and $\Im$ superscripts, respectively.} $\hat{H}_{FU}^\Im(z,z_1)$. Later, a phenomenological analysis of $A_N$ incorporated this twist-3 fragmentation piece along with the QS term~\cite{Kanazawa:2014dca}. The procedure employed by the authors of Ref.~\cite{Kanazawa:2014dca} was (1) to use a QCD equation of motion relation (EOMR) to write $H(z)$ in terms of $H_1^{\perp(1)}(z)$ and an integral of $\hat{H}_{FU}^\Im(z,z_1)$; (2) then fix $H_1^{\perp(1)}(z)$ in terms of the Collins function and $F_{FT}(x,x)$ in terms of the Sivers function, both of which had been previously extracted from TMD processes using a simple Gaussian ansatz
 for the respective transverse momentum dependence~\cite{Anselmino:2008sga,Anselmino:2013vqa}; (3) finally, fit $\hat{H}_{FU}^\Im(z,z_1)$ to the RHIC data.  The result was a successful description of $A_N$ data from RHIC for both neutral and charged pion production, where the twist-3 fragmentation piece dominated the asymmetry~\cite{Kanazawa:2014dca}.  This was a crucial step towards clarifying the mechanism behind TSSAs in hadronic collisions.  

Since that study, there has been a recent development, due to the work in Ref.~\cite{Kanazawa:2015ajw}, that has provided an additional operator constraint, a Lorentz invariance relation (LIR), on the twist-3 FFs $H_1^{\perp(1)}(z)$, $H(z)$, and an integral of
 $\hat{H}_{FU}^\Im(z,z_1)$ that enter $A_N$, which was not known at the time of the computation in Ref.~\cite{Kanazawa:2014dca}.  From a purely theoretical standpoint, the work in Ref.~\cite{Kanazawa:2015ajw} has elucidated that multi-parton (dynamical) correlators are the fundamental objects that cause transverse spin observables.  For example, one can solve the relevant LIR and EOMR and write $H_1^{\perp(1)}(z)$ and $H(z)$ in terms of integrals of $\hat{H}_{FU}^\Im(z,z_1)$.  That is, $\hat{H}_{FU}^\Im(z,z_1)$ is the ``base'' function from which $H_1^{\perp(1)}(z)$ and $H(z)$ are derived.  However, from a practical phenomenological standpoint, one can use LIRs to eliminate ``unknown'' functions in terms of ``known'' functions.  To be more specific, the only twist-3 functions that enter $A_N$ in $p^\uparrow  p\to \pi\,X$ that we {\it a priori} have any information on are the kinematical correlators because they are connected to TMD functions.\footnote{Of course there are still uncertainties in these TMD inputs because they are not well-constrained in the forward $x_F$ region where $A_N$ is large.}  
 Therefore, we can use these relations to re-express $A_N$ in terms of the maximum number of kinematical correlators.\footnote{Note from our discussion before that we can never write the entire cross section only in terms of kinematical functions.}

To this end, we present a new analysis of the {\rm TSSA} in $p^\uparrow  p\to \pi\,X$, where we compute the QS term and the twist-3 fragmentation piece
as was done in~\cite{Kanazawa:2014dca}.  However, we do not aim at a fit of the 3-parton FF $\hat{H}_{FU}^\Im(z,z_1)$.  Rather, we first employ the LIR and EOMR to rewrite the cross section, eliminating unknown functions where possible.  We then determine the contribution only from the kinematical correlators, where we use the latest TMD evolved extractions of those functions.  Consequently, we are able to provide a constraint on the parts that remain from the lesser known dynamical functions. The Letter is organized as follows: in Sec.~\ref{s:analytical} we review the relevant analytical formulae needed in our analysis, and, in particular, rewrite the fragmentation term using the aforementioned EOMR and LIR. Next, in Sec.~\ref{s:phenom} we conduct our numerical study of $A_N$ and compare our results to experimental data. Finally, in Sec.~\ref{s:sum} we summarize our work.

\section{The Qiu-Sterman and fragmentation contributions to  
$p^\uparrow p\to \pi\, X$ \label{s:analytical}}
We consider TSSAs in the single-inclusive production of pions from proton-proton collisions, 
\begin{equation}
p(P, S_{T}) + p(P') \rightarrow \pi(P_h) + X\,, 
\end{equation}
where we have indicated the momenta and polarizations of the particles. 
The spin-averaged differential cross section $d\sigma$ in Eq.~(\ref{e:AN}) at leading-order can be written as
\begin{align}
E_h\frac{d\sigma} {d^{3}\vec{P}_{h}} = \frac{\alpha_S^2} {S} \sum_i \sum_{a,b,c}
\int_0^1\!\frac{dz} {z^2}\int_0^1 \!\dfrac{dx'} {x'}\int_0^1 \!\dfrac{dx} {x}\,\,\delta(\hat{s}+\hat{t}+\hat{u})
\,f_1^a(x)\,f_1^b(x')\,D_1^{\pi/c}(z)\,S_U^i\,,
\label{e:spin-avg}
\end{align}
where $\sum_i$ is a sum over all partonic interaction channels, $a$ can be a quark, anti-quark, or gluon and likewise for $b,c$, $\alpha_s$ is the strong coupling constant, and $f_1(x)$ ($D_1(z)$) is the standard twist-2 unpolarized PDF (FF). We have made explicit that parton $c$ fragments into a pion. The well-known hard factors  for the unpolarized cross section are denoted
by $S_U^i$~\cite{Owens:1986mp,Kang:2013ufa} and can be found in, e.g., Appendix A of Ref.~\cite{Kouvaris:2006zy}. They are functions of the partonic Mandelstam variables $\hat{s} = xx' S,\,\hat{t} = xT/z,\,{\rm and}\;\hat{u} = x'U/z$, where $S = (P+P')^2$, $T = (P-P_h)^2$, and $U = (P'-P_h)^2$.

Let us now turn to the spin-dependent differential cross section $d\Delta\sigma(S_T)$ in Eq.~\eqref{e:AN}. All three terms in Eq.~(\ref{e:collfac}) enter into the analysis. However, as stated in Section~\ref{sec:intro}, we will focus on the $qgq$ SGP (QS) piece of the first term and the third (fragmentation) term: 
\begin{align}
E_h\frac{d\Delta\sigma(S_T)} {d^{3}\vec{P}_{h}} = E_h\frac{d\Delta\sigma^{QS}\!(S_T)} {d^{3}\vec{P}_{h}}
+ E_h\frac{d\Delta\sigma^{Frag}\!(S_T)} {d^{3}\vec{P}_{h}}.
\label{e:spin-dep}
\end{align}
The definitions of the relevant functions can be found in Refs.~\cite{Pitonyak:2016hqh,Kanazawa:2015ajw}.  
First, we give the expression for the QS term, which reads~\cite{Qiu:1998ia,Kouvaris:2006zy}
\begin{align}
E_h\frac{d\Delta\sigma^{QS}\!(S_T)} {d^{3}\vec{P}_{h}}
=& -\frac{4\alpha_S^2 M} {S}\,\epsilon^{P'\!PP_h S_T}\sum_i\sum_{a,b,c}\int_0^1\!\frac{dz} {z^3}\int_0^1 \!dx'\int_0^1 \!dx\,\,\delta(\hat{s}+\hat{t}+\hat{u})\frac{\pi} {\hat{s}\hat{u}} 
\nonumber\\
&\times\,f_1^b(x')\,D_1^{\pi/c}(z)\left[F_{FT}^a(x,x)-x\frac{dF_{FT}^a(x,x)} {dx}\right]S^i_{F_{FT}}\,,
\label{finalcr}
\end{align}
where the Levi-Civita tensor is defined with $\epsilon^{0123} = +1$, and the hard factors are denoted by $S_{F_{FT}}^i$ and can be found in Appendix A of Ref.~\cite{Kouvaris:2006zy}. 
There is an operator identity that relates the QS function $F^q_{FT}(x,x)$ to the first $k_T$-moment of the Sivers function~\cite{Boer:2003cm},
\begin{equation}
 \pi F^q_{FT}(x,x) =  f_{1T}^{\perp(1),q}(x)\big|_{\rm SIDIS} = -f_{1T}^{\perp(1),q}(x)\big|_{\rm DY}\, , 
 \label{e:QS_Siv}
\end{equation}
where
\begin{equation}
f_{1T}^{\perp(1),q}(x)\equiv\!\int\!d^2\vec{k}_T \,\frac{\vec{k}_T^2} {2M^2} f_{1T}^{\perp, q}(x,\vec{k}_T^2)\,.
\label{e:first_mom}
\end{equation}
In Eq.~(\ref{e:QS_Siv}) we have indicated that the Sivers function is either the one extracted from semi-inclusive deep-inelastic scattering (SIDIS) or the Drell-Yan (DY)  process~\cite{Brodsky:2002cx, Collins:2002kn}.

Next, we look at the fragmentation term, which was first fully calculated in Ref.~\cite{Metz:2012ct} and reads 
\begin{align}
E_h\frac{d\Delta\sigma^{Frag}(S_T)} {d^{3}\vec{P}_{h}} =& -\frac{4\alpha_{s}^{2}M_{h}} {S}\, \epsilon^{P'\!PP_h S_T}\sum_{i}\sum_{a,b,c}\int_{0}^{1}\frac{dz} {z^{3}} \int_{0}^{1}\!dx' \int_0^1 \!dx\,\,\delta(\hat{s}+\hat{t}+\hat{u})
\frac{1} {\hat{s}\,(-x'\hat{t}-x\hat{u})}\,\nonumber\\ 
& \hspace{-2.0cm}\times
 \,h_{1}^{a}(x)\,f_{1}^{b}(x')\left\{\left[H_1^{\perp(1),\pi/c}(z)-z\frac{dH_1^{\perp(1),\pi/c}(z)} {dz}\right]S_{H_1^{\perp}}^{i} + \frac{1} {z} H^{\pi/c}(z)\, S_{H}^{i}\  \right. 
 + \frac{2} {z}\int_z^\infty\! \frac{dz_1} {z_1^2}\frac{1} {\left(\frac{1} {z}-\frac{1} {z_{1}}\right)^{\!2}}\, \hat{H}_{FU}^{\pi/c,\Im}(z,z_{1}) \,S_{\hat{H}_{FU}}^{i}\Bigg\}\,,
\label{e:sigmaFrag}
\end{align}
where $M_h$ is the pion mass, and $h_1(x)$ is the standard twist-2 transversity PDF. The functions $H_1^{\perp(1)}(z)$, $H(z)$, and $\hat{H}_{FU}^\Im(z,z_1)$ are, respectively, the kinematical, intrinsic, and dynamical unpolarized twist-3 FFs discussed in Section \ref{sec:intro}. The hard factors associated with them are represented by $S^i$ with the corresponding subscript, and they can be found in Appendix A of Ref.~\cite{Metz:2012ct}.\footnote{Note that in Ref.~\cite{Metz:2012ct}, $\hat{H}(z)\equiv H_1^{\perp(1)}(z)$.} Note that $H_1^{\perp(1)}(z)$ is the first $p_\perp$-moment of the Collins function, 
\begin{equation}
H_1^{\perp(1),q}(z) \equiv z^2\int \!d^2 \vec{p}_{\perp} \, \frac{\vec{p}_{\perp}^{\,2}}{2 M_h^2} \, 
H_{1}^{\perp,q}(z,z^2\vec{p}_{\perp}^{\,2})\,.
 \label{e:H1perp}
\end{equation}
These collinear twist-3 FFs are related to each other through an EOMR~\cite{Politzer:1980me,Boer:1997bw,Bacchetta:2006tn,Kang:2010zzb,Metz:2012ct,Kanazawa:2013uia},
\begin{equation}
H^{q}(z) \, = \, -2 z\, H_1^{\perp(1),q}(z) 
           + 2 z \, \int_{z}^{\infty} \frac{dz_1} {z_1^2} \,\, \frac{1} {\frac{1} {z}-\frac{1} {z_{1}}} \hat{H}_{FU}^{q,\Im}(z,z_{1}) \,.
\label{e:EOM_FF}
\end{equation}
It is important to mention that there is a similar EOMR involving TMD FFs~\cite{Boer:1997mf,Bacchetta:2006tn},
\bea
H^q(z,z^2\vec{p}_\perp^{\,2})=-\frac{z\vec{p}_\perp^{\,2}}{M_h^2}H_1^{\perp, q}(z,z^2\vec{p}_\perp^{\,2})+ \tilde{H}^{q}(z,z^2\vec{p}_\perp^{\,2})\,,
\label{e:EOMKT}
\eea
where $H(z,z^2\vec{p}_\perp^{\,2})$ is the TMD version of $H(z)$, $H_1^\perp(z,z^2\vec{p}_\perp^{\,2})$ is the usual TMD Collins function, and $\tilde{H}(z,z^2\vec{p}_\perp^{\,2})$ is a twist-3 $qgq$ TMD FF.
Upon integration over $\vec{p}_\perp$, we find\footnote{Note that $D(z)=z^2\int d^2\vec{p}_\perp D(z,z^2\vec{p}_\perp^{\,2})$ for a generic FF.}
\begin{equation}
H^{q}(z) \, = \, -2 z\, H_1^{\perp(1),q}(z) 
           + \tilde{H}^q(z)\,.
\label{e:EOM_FF2}
\end{equation}
Therefore, comparing Eqs.~\eqref{e:EOM_FF} and \eqref{e:EOM_FF2} one can make the identification\footnote{This connection was made previously in Ref.~\cite{Lu:2015wja}, where $H(z,z^2\vec{p}_\perp^{\,2})$ and $\tilde{H}(z,z^2\vec{p}_\perp^{\,2})$ were calculated in a spectator model, and in Ref.~\cite{Wang:2016tix} in the context of the asymmetry $A_{UT}^{\sin\phi_S}$ in SIDIS.}  (using the variable substitution $y=z/z_1$)
\bea
\tilde{H}^q(z) = 2z\int_0^1 \!dy\,\frac{1}{1-y}
\hat{H}^{q,\Im}_{FU}(z,z/y)\,. \label{e:Htilde}
\eea
That is, the r.h.s.~of Eq.~(\ref{e:Htilde}) is actually a TMD FF integrated over $\vec{p}_\perp$.  We will revisit this observation below. 

Another set of formulae relating the intrinsic, kinematical, and dynamical functions are LIRs~\cite{Kodaira:1998jn,Belitsky:1997ay,Eguchi:2006qz,Accardi:2009au,Kanazawa:2015ajw}.  For twist-3 FFs, these expressions were derived for the first time in Ref.~\cite{Kanazawa:2015ajw} using identities among non-local
operators and taking into account constraints from Lorentz invariance. For our purposes, we focus on the one relating $H_1^{\perp(1)}(z),\, H(z),\,{\rm and}\; \hat{H}_{FU}^\Im(z,z_1)$~\cite{Kanazawa:2015ajw}:
\begin{align}
  \frac{H^q(z)}{z} &= - \left( 1 - z \frac{d}{dz} \right) H_{1}^{\perp
  (1),q}(z)  - \frac{2}{z} \int_z^\infty \frac{dz_1}{z_1^2}
  \frac{\hat{H}_{FU}^{q,\Im}(z,z_1)}{(1/z-1/z_1)^2}\,. \label{LIRH}
\end{align}
We now employ both the EOMR (\ref{e:EOM_FF}) and LIR (\ref{LIRH})
to re-express the fragmentation contribution to $A_N$ in Eq.~\eqref{e:sigmaFrag} in terms
of  the kinematical function $H_{1}^{\perp (1)}(z)$
and the the dynamical correlator $\hat{H}_{FU}^{\Im}(z,z_1)$ (via $\tilde{H}(z)$), 
\begin{align}
E_h\frac{d\Delta\sigma^{Frag}(S_T)} {d^{3}\vec{P}_{h}} =& -\frac{4\alpha_{s}^{2}M_{h}} {S}\, \epsilon^{P'\!PP_h S_T}\sum_{i}\sum_{a,b,c}\int_{0}^{1}\frac{dz} {z^{3}} \int_{0}^{1}\!dx' \int_0^1 \!dx\,\,\delta(\hat{s}+\hat{t}+\hat{u}) \frac{1} {\hat{s}}\,\nonumber\\ 
&\times\,h_{1}^{a}(x)\,f_{1}^{b}(x')\left\{\left[H_1^{\perp(1),\pi/c}(z)-z\frac{dH_1^{\perp(1),\pi/c}(z)} {dz}\right]\tilde{S}_{H_1^{\perp}}^{i}
 +\left[-2H_{1}^{\perp(1),\pi/c}(z) + \frac{1} {z}\tilde{H}^{\pi/c}(z)\right] \tilde{S}_{H}^{i}\right\}\,,  
\label{e:DsigmaFragg}
\end{align}
where we have used the identity \eqref{e:Htilde} for the last term in brackets.    The hard factors are given by
\begin{align}
\tilde S_{\!H_{1}^{\perp}}^i\equiv\frac{S_{\!H_{1}^{\perp}}^i-S_{\!H_{FU}}^i}{-x'\hat t-x\hat u}\quad {\rm and} \quad \tilde S_{\!H}^i\equiv\frac{S_{\!H}^i-S_{\!H_{FU}}^i}{-x'\hat t-x\hat u}, \label{e:Stilde}
\end{align}
  which explicitly read
\begin{align}
&\tilde S_{\!H_{1}^{\perp}}^{q g\to q g} 
=-\frac{1}{N_c^2}\frac{1}{\tht}
+\nctmo\frac{\ths(\thu-\ths)}{\tht^3}
-\frac{\ths^2}{\tht^2\thu}, \quad
\tilde S_{\!H}^{qg\to qg}=
\frac{1}{N_c^2-1}\frac{\ths(\thu-\ths)}{\tht^3}
+\frac{1}{N_c^2}\frac{\ths-\thu}{2\tht\thu}
+\frac{(\ths-\thu)(\tht^2-2\tht\thu-2\thu^2)}{2\tht^3\thu}, 
\label{eq:sinitial}\\
&\tilde S_{\!H_{1}^{\perp}}^{qq'\to qq'}=
\frac{1}{N_c^2}\frac{\ths(\thu-2\tht)}{\tht^3}+\frac{\ths}{\tht^2}\,,\quad
\tilde S_{\!H}^{qq'\to qq'}=
\frac{1}{N_c^2}\frac{\ths(2\thu-\tht)}{\tht^3}-\frac{\ths\thu}{\tht^3}\,,\\
&\tilde S_{\!H_{1}^{\perp}}^{qq\to qq}=
\frac{1}{N_c^3}\frac{\ths(\tht-\thu)}{\tht^2\thu}
+\frac{1}{N_c^2}\frac{\ths(\thu-2\tht)}{\tht^3}
+\frac{\ths}{\tht^2}\,,\quad
\tilde S_{\!H}^{qq\to qq}=
\frac{1}{N_c^3}\frac{\ths(\tht-3\thu)}{2\tht^2\thu}
+\frac{1}{N_c^2}\frac{\ths(2\thu-\tht)}{\tht^3}
-\frac{1}{N_c}\frac{\ths^2}{2\tht^2\thu}
-\frac{\ths\thu}{\tht^3}\,,\\
&\tilde S_{\!H_{1}^{\perp}}^{q\bar q\to q\bar q}=
\frac{1}{N_c^3}\frac{\ths}{\tht^2}
+\frac{1}{N_c^2}\frac{\ths(\tht-\ths)}{\tht^3}
-\frac{1}{N_c}\frac{1}{\tht}\,,\quad
\tilde S_{\!H}^{q\bar q\to q\bar q}=
\frac{1}{N_c^3}\frac{3\ths-\tht}{2\tht^2}
+\frac{1}{N_c^2}\frac{\ths(\tht-2\ths)}{\tht^3}
+\frac{1}{N_c}\frac{\thu}{2\tht^2}
+\frac{\ths^2}{\tht^3}\,,\\
&\tilde S_{\!H_{1}^{\perp}}^{\bar q q\to q\bar q}=
\frac{1}{N_c^3}\frac{\ths}{\tht\thu}
-\frac{1}{N_c}\frac{1}{\tht}\,,\quad
 \tilde S_{\!H}^{\bar qq\to q\bar q}=
\frac{N_c^2+1}{N_c^3}\frac{\ths-\thu}{2\tht\thu}\,,\\
&\tilde S_{\!H_{1}^{\perp}}^{q\bar q'\to q\bar q'}=\frac{1}{N_c^2}\frac{\ths(\tht-\ths)}{\tht^3}\,,\quad
\tilde S_{\!H}^{q\bar q'\to q\bar q'}=
\frac{1}{N_c^2}\frac{\ths(\tht-2\ths)}{\tht^3}
+\frac{\ths^2}{\tht^3}\,.
\label{eq:sfinal}
\end{align}
Note that the factor $1/(-x'\hat{t}-x\hat{u})$ in Eq.~(\ref{e:Stilde}) is cancelled in each of the above expressions (\ref{eq:sinitial})--(\ref{eq:sfinal}).  Such a simplification also occurred for $A_N$ in $\ell \, p^\uparrow\to\pi\,X$ after using the LIR (\ref{LIRH})~\cite{Kanazawa:2015ajw} and appears to be a non-trivial cross-check of the result.
We also mention that one can obtain the channels involving anti-quark fragmentation by charge conjugating the above partonic processes $ab\rightarrow cd$.  The hard parts are the same as the ones above, i.e., $\tilde{S}^{\!\bar{a}\bar{b}\to\bar{c}\bar{d}} = \tilde{S}^{ab\to cd}$.

At this point, we  emphasize again that $\tilde{H}(z)$ in Eq.~(\ref{e:DsigmaFragg}) is the twist-3 $qgq$ TMD FF $\tilde{H}(z,z^2\vec{p}_\perp^{\,2})$ integrated over $\vec{p}_{\perp}$. Thus, we stress   
that one could obtain some information on $\hat{H}_{FU}^\Im(z,z_1)$ (via $\tilde{H}(z)$ or $\tilde{H}(z,z^2\vec{p}_\perp^{\,2})$) through transverse spin observables within the collinear twist-3 and/or TMD factorization formalisms. For example, the processes $p^\uparrow p\to \pi X$ and $\ell \, p^\uparrow\to\pi\,X$ depend on $\tilde{H}(z)$. On the other hand, at small-$q_T$, where $q$ is the momentum of the virtual photon, the unintegrated correlator $\tilde{H}(z,z^2\vec{p}_\perp^{\,2})$ enters (along with several other terms) the asymmetries $A_{UT}^{\sin\phi_S}$, $A_{UU}^{\cos\phi_h}$, $A_{UL}^{\sin\phi_h}$, and $A_{UT}^{\sin(2\phi_h-\phi_S)}$ in SIDIS~\cite{Bacchetta:2006tn} as well as the analogous ones in $e^+e^-\to h_a\,h_b\,X$~\cite{Boer:1997mf}. 
 At the same time, a measurement of $A_{UT}^{\sin\phi_S}$ in SIDIS integrated over $P_{hT}$ is sensitive to a single term\footnote{In $e^+e^-\to h_a\,h_b\,X$ there is an additional term coupling $D_T(z_a)$ to $D_1(z_b)$, whereas in SIDIS the analogous term coupling $f_T(x)$ to $D_1(z)$ vanishes due to time-reversal invariance.} that couples $\tilde{H}(z)$ to the transversity PDF $h_1(x)$~\cite{Bacchetta:2006tn}.\footnote{This corresponds to a leading-order (LO) computation for this asymmetry within the usual collinear twist-3 factorization formalism~\cite{Kang:2012ns}. At next-to-leading order (NLO), typically a more complicated structure will appear that likely contains more terms~\cite{Kanazawa:2013uia}.}   We want to reiterate once more that $\hat{H}_{FU}^\Im(z,z_1)$ (via $\tilde{H}(z)$ or $\tilde{H}(z,z^2\vec{p}_\perp^{\,2})$) does show up in several different observables  that can provide more information on this function in the future. In the next section, we perform an analysis of the TSSA for $p^\uparrow p\to \pi X$ and comment on what constraints on $\tilde{H}(z)$ it can provide. 


\section{Phenomenology \label{s:phenom}}

We now give our numerical estimate for $A_N$.  Using Eqs.~(\ref{e:spin-avg}), (\ref{finalcr}), (\ref{e:DsigmaFragg}),
we are able to express, respectively, the numerator and denominator of $A_N$ in Eq.~(\ref{e:AN}) as
\begin{align}
d\Delta \sigma(S_T) &= \frac{2P_{hT}\alpha_S^2}{S}  \displaystyle\sum_i\displaystyle\sum_{a,b,c}\!\displaystyle\int_{z_{min}}^1\!\dfrac{dz} {z^3}\displaystyle\int_{x_{min}}^1\!\dfrac{dx} {x}\dfrac{1} {x'}\dfrac{1}{xS+U/z} f_1^b(x') \left[M_h\,h_1^a(x)\,\mathcal{H}^{\pi/c,i}(x,x',z)+ \dfrac{M}{\hat{u}} \mathcal{F}^{a,i}(x,x',z)\,D_1^{\pi/c}(z)\right],\label{e:Num}\\[0.3cm]
d\sigma &= \frac{\alpha_S^2}{S} \displaystyle\sum_i\displaystyle\sum_{a,b,c}\displaystyle\int_{z_{min}}^1\!\dfrac{dz} {z^2}\displaystyle\int_{x_{min}}^1\!\dfrac{dx} {x}\dfrac{1} {x'}\dfrac{1} {xS+U/z}\,f_1^a(x)\,f_1^b(x')\,D_1^{\pi/c}(z)\,S_U^i\,, \label{e:Den}
\end{align}
where $z_{min}=-(T+U)/S$, $x_{min} = -(U/z)/(T/z+S)$, and $x' = -(xT/z)/(xS+U/z)$. 
In Eq.~(\ref{e:Num}) the quantities $\mathcal{H}^{\pi/c,i}(x,x',z)$ and $\mathcal{F}^{a,i}(x,x',z)$ are
\begin{align}
\mathcal{H}^{\pi/c,i}(x,x',z)& = \left[H_1^{\perp(1),\pi/c}(z)-z\frac{dH_1^{\perp(1),\pi/c}(z)} {dz}\right]\tilde{S}_{H_1^{\perp}}^{i}
 +\left[-2H_{1}^{\perp(1),\pi/c}(z) + \frac{1} {z}\tilde{H}^{\pi/c}(z)\right] \tilde{S}_{H}^{i}\,, \label{e:mathcalH} \\[0.3cm]
 \mathcal{F}^{a,i}(x,x',z) &=  \pi\left[ F_{FT}^a(x,x) -x\frac{d F_{FT}^a(x,x)} {dx}\right]S^i_{F_{FT}}\,. 
\label{e:mathcalF}
\end{align}
We focus on the contributions in Eqs.~(\ref{e:mathcalH}), (\ref{e:mathcalF}) from the functions $ F_{FT}(x,x)$ and $H_1^{\perp(1)}(z)$, for which we have information on from the Sivers function and Collins function, respectively, that have been extracted from TMD processes~\cite{Anselmino:2007fs,Anselmino:2008jk,Anselmino:2008sga,Anselmino:2012aa,Anselmino:2013vqa,Anselmino:2013rya,Echevarria:2014xaa,Anselmino:2015sxa,Kang:2015msa}.  
In  our  analysis  we will ignore $
\tilde{H}(z)$ in Eq.~\eqref{e:mathcalH}, which is 
 equivalent to approximating Eq.~(\ref{e:EOM_FF2}) as
\begin{equation}
H^{q}(z) \, = \, -2 z\, H_1^{\perp(1),q}(z) 
           + \tilde{H}^q(z)\, \approx -2 z\, H_1^{\perp(1),q}(z) \; .
\label{e:EOM_FF2_WW}
\end{equation}
This is known as a Wandzura-Wilczek-type approximation \cite{Wandzura:1977qf}, which for TMDs was explored in Refs.~\cite{Metz:2008ib,Teckentrup:2009tk}.  We emphasize that Eq.~(\ref{e:EOM_FF2_WW}) is a statement that $\tilde{H}(z)/z$ is parametrically smaller than $H_1^{\perp(1)}(z)$, {\it not} that $\hat{H}^\Im(z,z_1)$ is zero. In fact, $\hat{H}_{FU}^\Im(z,z_1)$ {\it must be nonzero} because it was shown in Ref.~\cite{Kanazawa:2015ajw} that $H_1^{\perp(1)}(z)$ is an integral of $\hat{H}_{FU}^\Im(z,z_1)$ .  That is, $\hat{H}_{FU}^\Im(z,z_1) = 0$ implies $H_1^{\perp(1)}(z) = 0$, and consequently, Eq.~(\ref{e:mathcalH}) would vanish identically. Moreover, we know from current extractions of the Collins function that $H_1^{\perp(1)}(z) \neq 0$. 
Here, the purpose of our computation is not to offer a complete analysis of $A_N$; it is to use  recent TMD evolved extractions of known (i.e., kinematical) inputs to the observable, along with a new constraint from the LIR~\eqref{LIRH}, to assess how well we are currently able to describe the data and ascertain what contributions remain from the dynamical functions. This will help guide a future fit of these correlators, in particular $\hat{H}_{FU}^\Im(z,z_1)$ (or $\tilde{H}(z)$), where one would be able to confirm or refute the approximation (\ref{e:EOM_FF2_WW}).  The function $\hat{H}_{FU}^\Im(z,z_1)$ was originally extracted in Ref.~\cite{Kanazawa:2014dca} before the LIR (\ref{LIRH}) was derived, and, therefore, that work must be updated to include this constraint.

\begin{figure}
\centering
\includegraphics[width=7.cm]{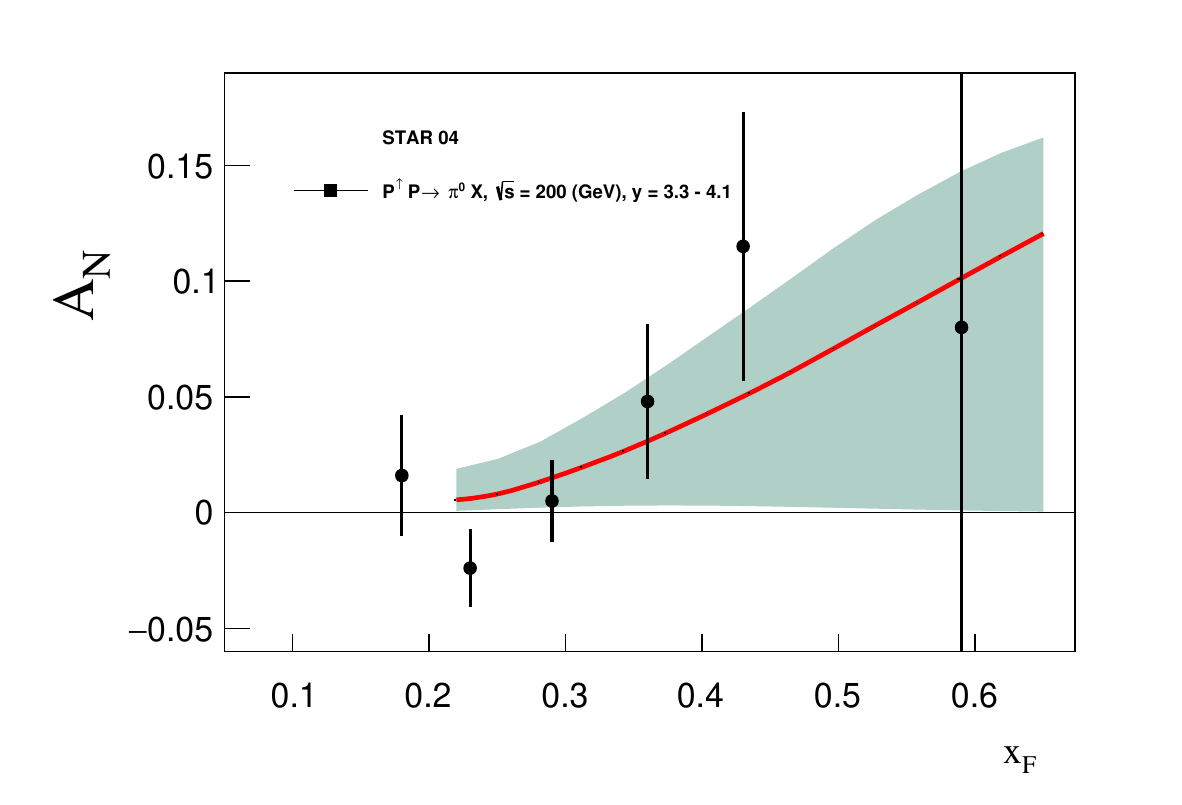}
\includegraphics[width=7.cm]{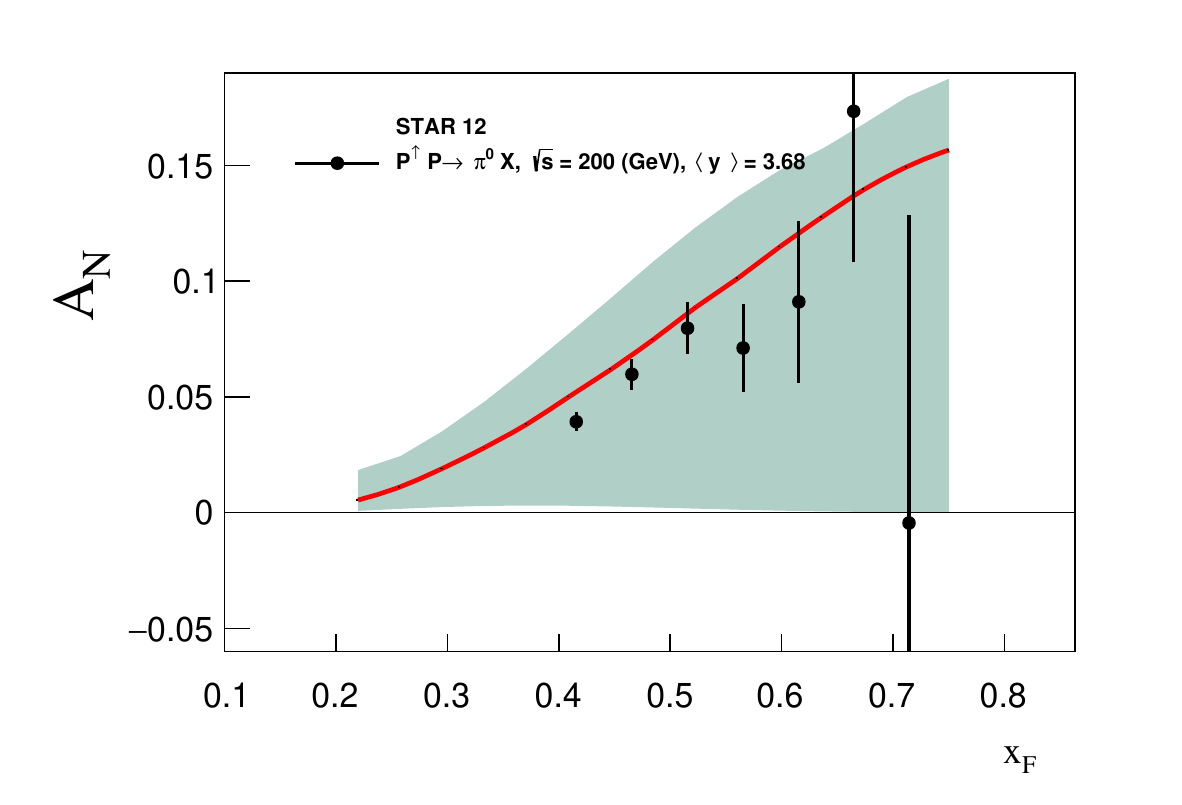}
\\ \vskip -0.3cm \hskip 0.5cm \tiny (a)\hskip 7cm (b) \\
\includegraphics[width=7.cm]{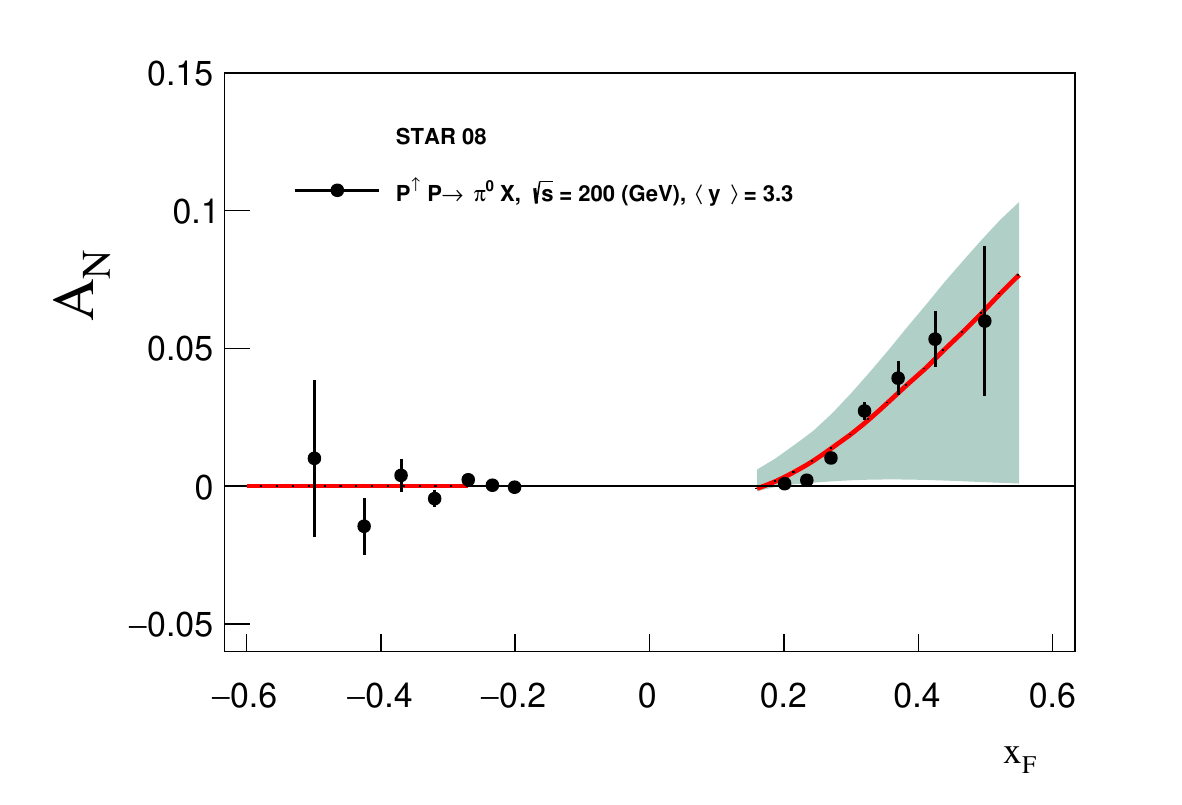}
\includegraphics[width=7.cm]{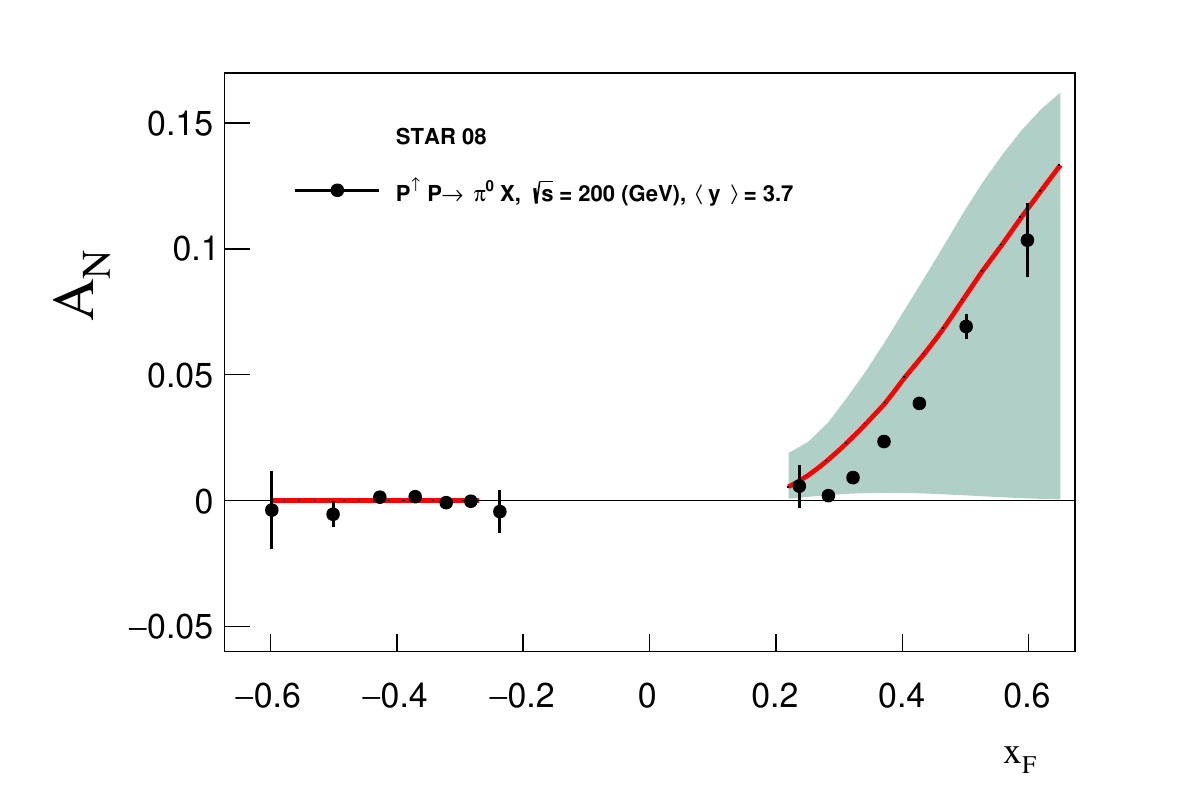}
\\ \vskip -0.3cm \hskip 0.5cm (c)\hskip 7cm (d) \\
\includegraphics[width=7.cm]{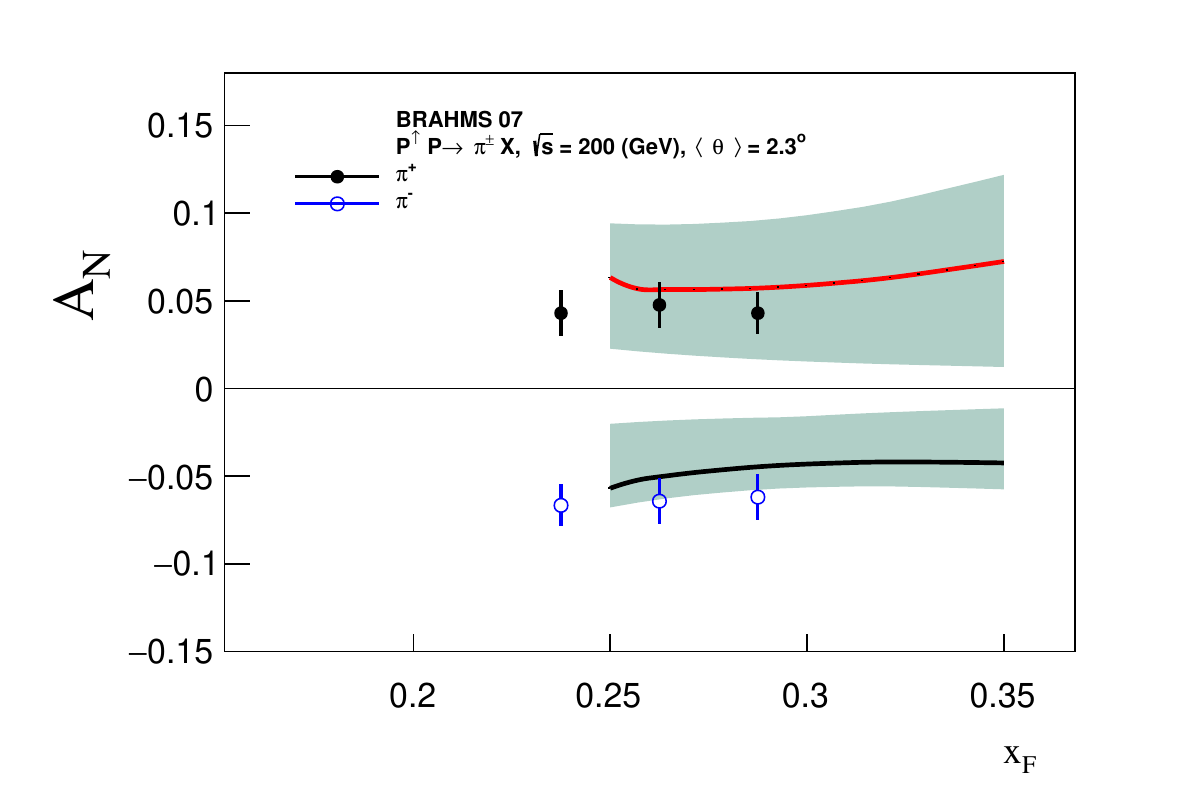}
\includegraphics[width=7.cm]{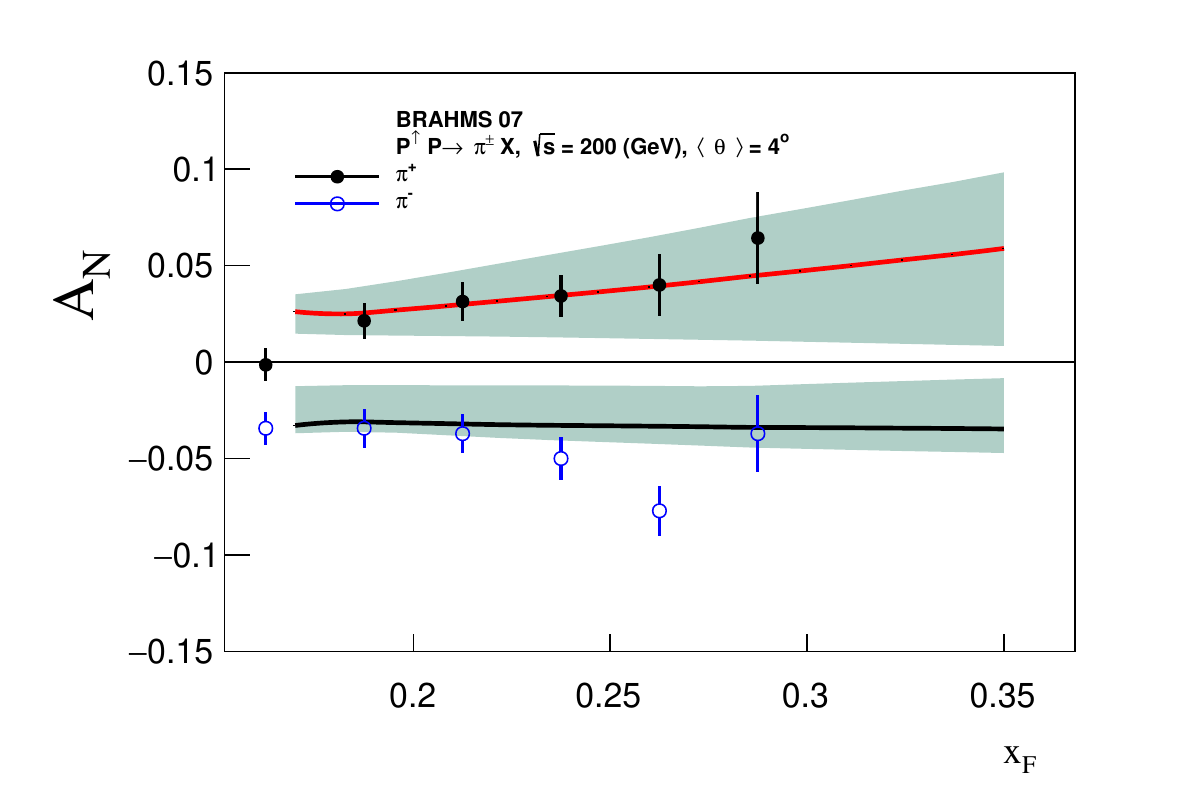}
\\ \vskip -0.3cm \hskip 0.5cm (e)\hskip 7cm (f)
\caption{Calculations of the contribution to $A_N$ from Eq.~(\ref{e:DsigmaFragg}) (using the approximation (\ref{e:EOM_FF2_WW})) compared to
(a) STAR Collaboration 2004 data ~\cite{Adams:2003fx} on $A_N$ for $\pi^0$, 
(b) STAR Collaboration 2012 data ~\cite{Abelev:2008af} on $A_N$ for $\pi^0$  at $\langle y \rangle = 3.68$, 
(c) STAR Collaboration 2008 data ~\cite{Adamczyk:2012xd} on $A_N$ for $\pi^0$  at $\langle y \rangle = 3.3$,   
(d) STAR Collaboration 2008 data ~\cite{Adamczyk:2012xd} on $A_N$ for $\pi^0$  at $\langle y \rangle = 3.7$,  
(e)  BRAHMS Collaboration 2007 data ~\cite{Lee:2007zzh} on $A_N$ for $\pi^\pm$ (black closed data $\pi^+$, blue open data $\pi^-$)   at $\langle \theta \rangle = 2.3^o$,    
(f)  BRAHMS Collaboration 2007 data ~\cite{Lee:2007zzh} on $A_N$ for $\pi^\pm$ (black closed data $\pi^+$, blue open data $\pi^-$)   at $\langle \theta \rangle = 4^o$.    
The solid lines correspond to the results using the central parameters from Ref.~\cite{Kang:2015msa} for $h_1(x)$ and $H_1^{\perp(1)}(z)$. 
The shaded regions correspond to an estimate of 90\% C.L. error band from Ref.~\cite{Kang:2015msa} due to uncertainties in $h_1(x)$ and $H_1^{\perp(1)}(z)$.}  
\label{fig:star_pi0}  
\end{figure}

As mentioned above, we will ignore $\tilde{H}(z)$ and compute the terms in Eqs.~(\ref{e:mathcalH}), (\ref{e:mathcalF}) that involve $F_{FT}(x,x)$ and $H_1^{\perp(1)}(z)$, using the latest fits of  the Sivers and Collins functions that incorporate TMD evolution. 
We elaborate a bit here. TMD evolution is conventionally carried out in  coordinate space, and thus one usually defines the Fourier transform of the TMD in a two-dimensional $\vec{b}$-space~\cite{Collins:2011zzd}. Take the quark Sivers function as an example~\cite{Echevarria:2014xaa,Aybat:2011ge},  
\begin{align}
f_{1T}^{\perp, q(\alpha)}(x, \vec{b}; Q) = \frac{1}{M} \int \!d^2\vec{k}_T e^{-i\vec{k}_T\cdot \vec{b}} \,\vec{k}_T^{\alpha}\, f_{1T}^{\perp, q}(x, \vec{k}_T^2; Q),
\end{align}
which is probed at the momentum scale $Q$. TMD evolution evolves the above Sivers function from an initial low scale $Q_0$ to a high $Q$: 
$f_{1T}^{\perp, q(\alpha)}(x, \vec{b}; Q_0) \rightarrow f_{1T}^{\perp, q(\alpha)}(x, \vec{b}; Q)$. At the same time, one can further perform an operator product expansion to relate the TMD at the initial scale $Q_0$ to a corresponding collinear function~\cite{Aybat:2011ge}. For the Sivers function at small $b$, one has a connection to the QS function:
\begin{align}
f_{1T}^{\perp, q(\alpha)}(x, \vec{b}; Q_0) \propto \frac{i\vec{b}^\alpha}{2} \int_ x^1 \!\frac{dx'}{x'} C\left(x/x', b, Q_0\right)\otimes  F_{FT}(x',x', Q_0),
\end{align}
where the coefficient $C$ was derived in \cite{Kang:2011mr,Sun:2013hua}. Thus, in the global analysis of the Sivers asymmetry that incorporates TMD evolution, one can directly extract the QS function $F_{FT}(x,x)$. Similarly we can directly extract $H_1^{\perp(1)}(z)$ through an analysis of the Collins asymmetry with TMD evolution. For details, see Ref.~\cite{Kang:2015msa}.

In the following numerical calculations,  we use $H_1^{\perp(1)}(z)$ extracted in Ref.~\cite{Kang:2015msa}, evolved through the ``diagonal'' piece of its evolution equations~\cite{Yuan:2009dw,Kang:2010xv}. We found that the QS function extracted in Ref.~\cite{Echevarria:2014xaa} leads to a negligible contribution from Eq.~\eqref{e:mathcalF} to $A_N$, so we will ignore it in the plots below.  Instead, we focus on the fragmentation term in Eq.~\eqref{e:Num} driven by (\ref{e:mathcalH}), which gives the  dominant contribution to the asymmetry. This confirms the original findings in Ref.~\cite{Kanazawa:2014dca}.  A few additional comments are in order about the size of the QS term.  The smallness of this term is because of the large-$x$ behavior of the Sivers function~\cite{Anselmino:2008sga,Anselmino:2013vqa,Gamberg:2013kla}, which can be explored and constrained in future experiments, like those at the 12 GeV upgrade of Jefferson Lab (JLab 12)~\cite{Dudek:2012vr,Ye:2016prn}. More specifically, one can only generate a non-negligible QS term if the $\beta$ parameters (that control the large-$x$ behavior of the Sivers/QS function) are different for up and down quarks~\cite{Gamberg:2013kla,Kanazawa:2014dca}, whereas in Ref.~\cite{Echevarria:2014xaa} one has $\beta_u=\beta_d$.  While there does exist an extraction of the Sivers function where $\beta_u\neq\beta_d$~\cite{Anselmino:2013vqa}, this extraction is not at the NLL$'$ accuracy that we use for other functions in our analysis.  Moreover, if such a Sivers function is used, as explored in Ref.~\cite{Kanazawa:2014dca}, it  most likely will not overwhelm  the large effect from the fragmentation term. Nevertheless, it will be interesting to include both the QS functions and the fragmentation terms in the global analysis in order to understand $A_N$ fully. Such a study will be explored in the future. 

\begin{figure}[ht]
\centering
\includegraphics[width=8.cm]{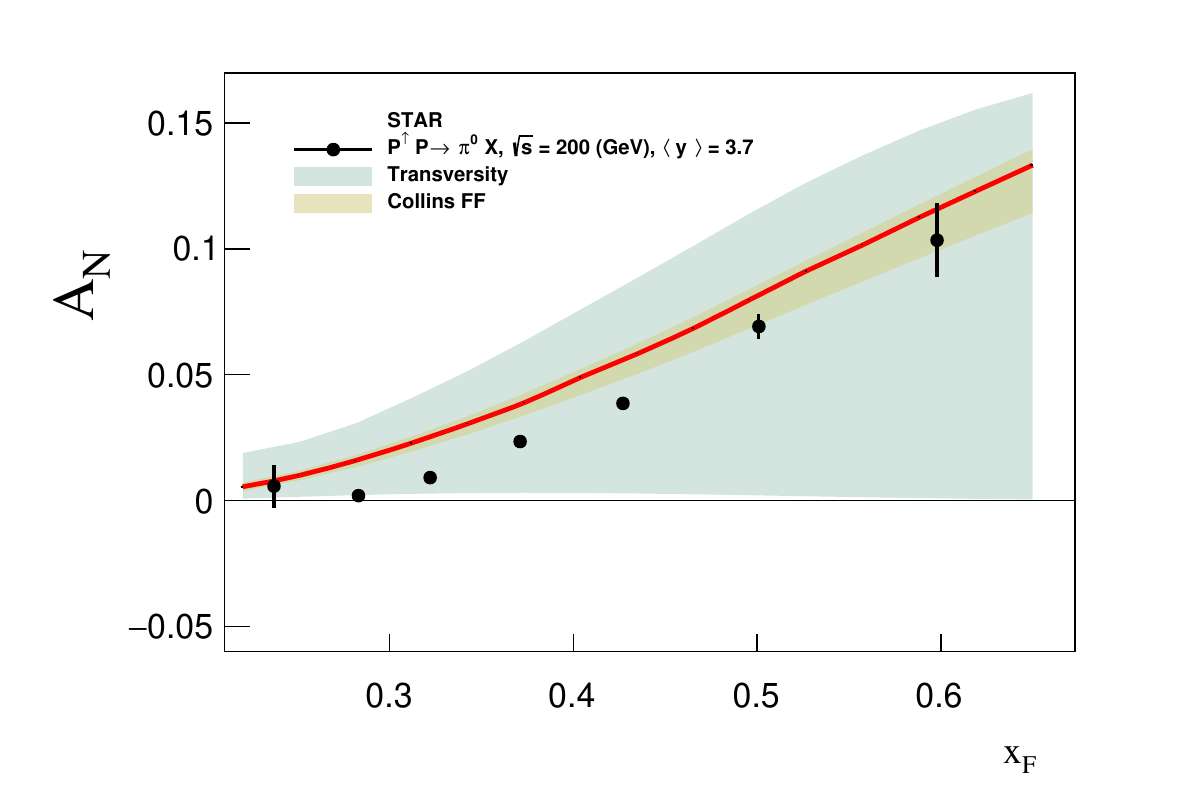} 
\includegraphics[width=8.cm]{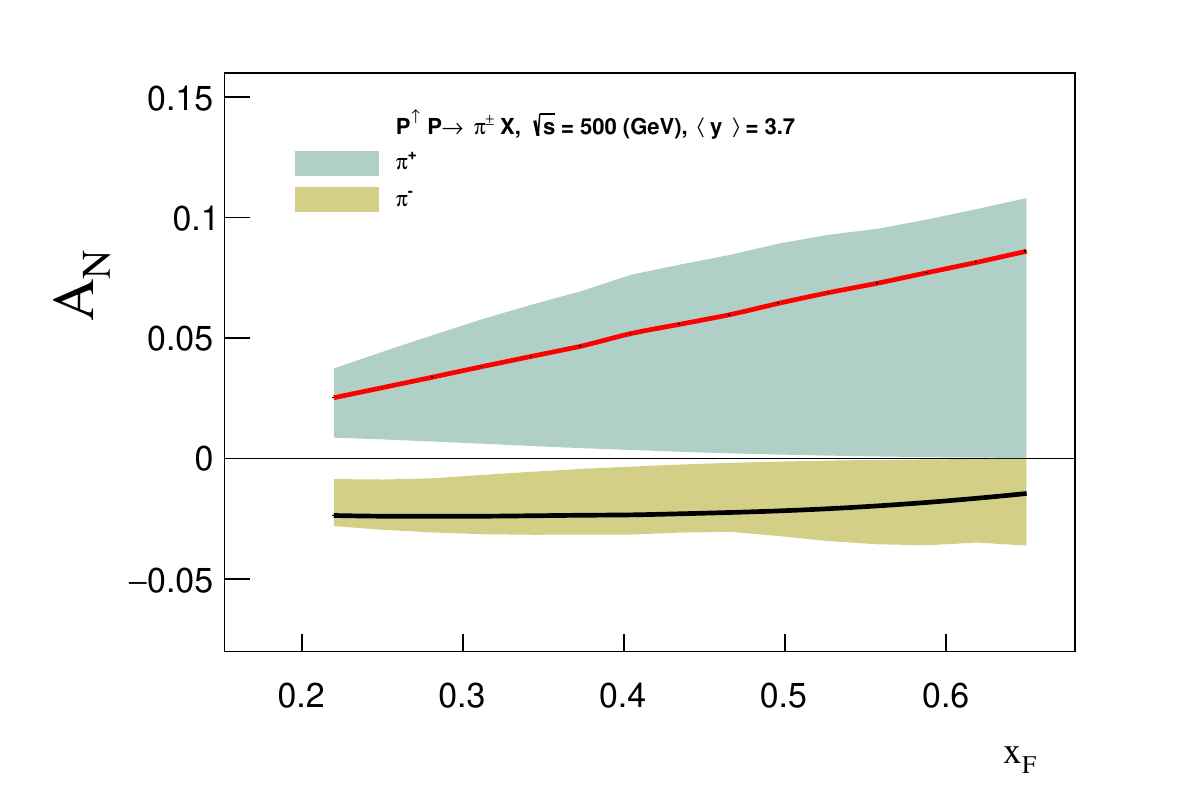} 
\\ \vskip -0.35cm \hskip 0.5cm \tiny (a)\hskip 8cm (b) \\
\caption{
(a) STAR Collaboration data on $A_N$ for $\pi^0$  as function of $x_F$ at $\langle y \rangle = 3.7$ for positive $x_F$. The two error bands correspond to the theoretical uncertainty from $H_1^{\perp(1)}(z)$ (the narrow band) and $h_1(x)$ (the wide band). 
(b) Predictions for  $A_N$ in $\pi^\pm$  production as function of $x_F$ at $\langle y \rangle = 3.7$ for positive $x_F$ using Eq.~(\ref{e:DsigmaFragg}) (along with the approximation (\ref{e:EOM_FF2_WW})). The error bands correspond to the theoretical uncertainty from both $h_1(x)$ and $H_1^{\perp(1)}(z)$.}  
\label{fig:star_pi0_error}
\end{figure}

In Fig.~\ref{fig:star_pi0}, we show the result of our calculation compared with the BRAHMS charged pion and STAR neutral pion data for $A_N$ as a function of $x_F$~\cite{Lee:2007zzh,Adams:2003fx,Abelev:2008af,Adamczyk:2012xd}. The error bands for $A_N$ in those plots are based on the uncertainties in $h_1(x)$ and $H_1^{\perp(1)}(z)$ from the fit in Ref.~\cite{Kang:2015msa}. In the large-$x_F$ (i.e., large-$x$ and large-$z$) region these functions are not well-constrained, and consequently, one obtains large errors due to these inputs. 
We see that, although the theoretical calculations undershoot or overshoot $A_N$ in some places, the central curves do a reasonable job in describing the data. We are especially encouraged by these plots given that the contribution from $\tilde{H}(z)$ still needs to be included. This clearly demonstrates that this function must be nonzero. Through this computation, we now have a constraint on $\tilde{H}(z)$ and leave a fit of this function to $A_N$ data for future work.  We emphasize again that the unintegrated  version of this correlator also enters multiple asymmetries in SIDIS and $e^+e^-\to h_a\,h_b\,X$, while $\tilde{H}(z)$ itself can be directly measured in $A_{UT}^{\sin\phi_S}$ in SIDIS integrated over $P_{hT}$.  

\begin{figure}[ht]
\centering
\includegraphics[width=8.cm]{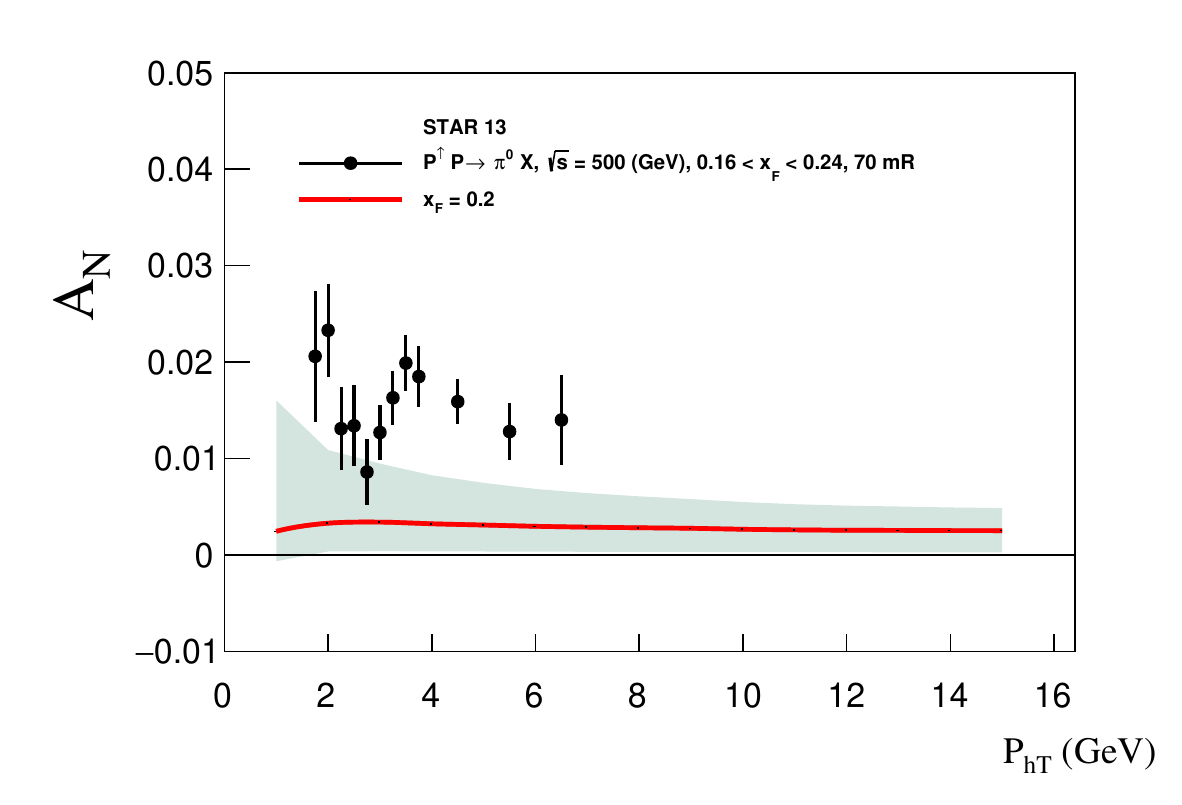}
\includegraphics[width=8.cm]{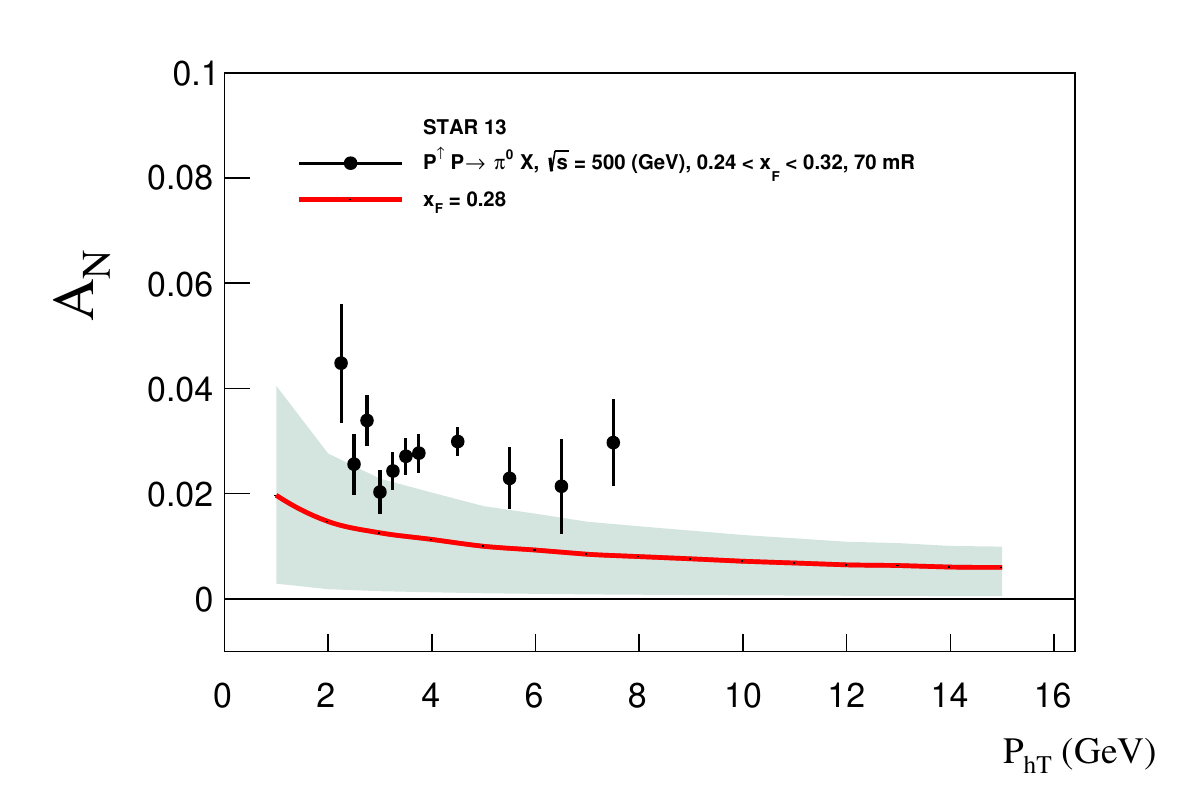}
\\ \vskip -0.3cm \hskip 0.5cm \tiny (a)\hskip 8cm (b) \\
\includegraphics[width=8.cm]{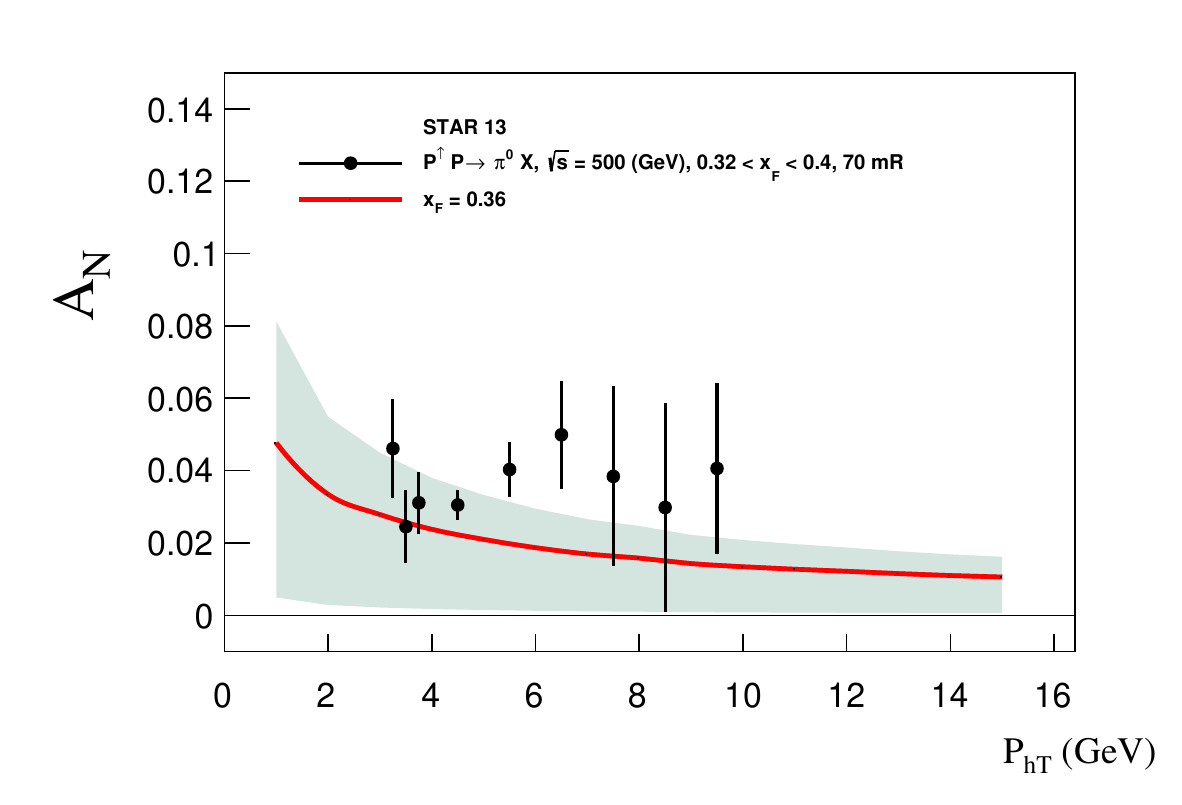}
\\ \vskip -0.3cm \hskip 0.5cm (c)  \\
 \caption{Calculations of the contribution to $A_N$ from Eq.~(\ref{e:DsigmaFragg}) (using the approximation (\ref{e:EOM_FF2_WW})) compared to
STAR Collaboration 2013 data ~\cite{Heppelmann:2013ewa} on $A_N$ for $\pi^0$ at $\sqrt{s} = 500$ GeV as a function of $P_{hT}$ (a)  $0.16 < x_F < 0.24$,
(b)  $0.24 < x_F < 0.32$, and (c)  $0.32 < x_F < 0.4$.
The solid lines correspond to results using the central parameters from Ref.~\cite{Kang:2015msa} for $h_1(x)$ and $H_1^{\perp(1)}(z)$. 
The shaded regions correspond to an estimate of 90\% C.L. error band from Ref.~\cite{Kang:2015msa} due to uncertainties in $h_1(x)$ and $H_1^{\perp(1)}(z)$.}  
\label{fig:star_pi0_pt}  
\end{figure}

Moreover, since $F_{FT}(x,x)$ and $H_1^{\perp(1)}(z)$, $h_1(x)$ enter the TMD evolution equations for the Sivers and Collins asymmetries, respectively, in SIDIS and $e^+e^-\to h_a\,h_b\,X$, one can eventually perform a global analysis that includes all these observables along with $A_N$ in proton-proton and lepton-nucleon collisions (where $F_{FT}(x,x)$, $\tilde{H}(z)$, $H_1^{\perp(1)}(z)$, $h_1(x)$ all enter).  This would better constrain the large-$x_F$ behavior of these functions and greatly reduce the error bands in our plots since we have data from RHIC in this region.
We found that the uncertainty in $h_1(x)$ in this regime is what dominates the error over that from $H_1^{\perp(1)}(z)$, see Fig.~\ref{fig:star_pi0_error}(a). 
Thus, it is evident that the $A_N$ data would allow us to drastically improve the extraction of transversity. Also, future measurements at JLab12 can improve the situation in the large-$x$ region~\cite{Ye:2016prn}. In order to demonstrate the powerful capability of RHIC future measurements~\cite{Aschenauer:2016our}, we present our predictions for $A_N$ in $\pi^\pm$ production at 500 GeV in Fig.~\ref{fig:star_pi0_error}(b). One can clearly see that large-$x_F$ measurements of $A_N$ will reduce the uncertainty of the large-$x$ behavior of transversity and, together with other data sets, allow us to explore the missing contribution from $\tilde{H}(z)$.
In addition, we also give our result for $A_N$ as a function of $P_{hT}$ in Fig.~\ref{fig:star_pi0_pt} compared with the STAR data from Ref.~\cite{Heppelmann:2013ewa}. One can see that our calculations exhibit a flat behavior, similar to that shown in Ref.~\cite{Kanazawa:2014dca}. The reason is that in the forward region, where $\hat t$  becomes very small, the $qg \to qg$ channel dominates, and the hard function $S_{\!H_{1}^{\perp}}^{qg \to qg} \propto 1/\hat t^3$ compensates the twist-3 $(P_{hT})^{-1}$ fall off of the asymmetry. Again, one has to keep in mind that there is still a term missing, $\tilde{H}(z)$ from our analysis, which needs to be fit to data as has been emphasized above. It is also important to emphasize  that the experimental data has a very large uncertainty which prevents an unambiguous identification of the $P_{hT}$-dependence. These open issues can only be addressed by future experimental measurements and theoretical work.

We end this section with a brief comment about the fragmentation contribution to $A_N$ in $p^\uparrow A\to \pi^0\,X$.  Recently, a calculation of this term was carried out in Ref.~\cite{Hatta:2016khv} that included gluon saturation effects in the unpolarized nucleus.  The authors found that the first two terms in braces in Eq.~(\ref{e:sigmaFrag}) are proportional to $A^{-1/3}$ (see also~\cite{Kang:2011ni}), while the third term is proportional to $A^0$. Since in Ref.~\cite{Kanazawa:2014dca} one finds that this third term is negligible (see Fig.~3 of Ref.~\cite{Kanazawa:2014dca}), the authors of Ref.~\cite{Hatta:2016khv} concluded that the fragmentation piece to $A_N$ in $p^\uparrow A$ collisions is proportional to $A^{-1/3}$, which is in contradiction to recent STAR measurements~\cite{Heppelmann:SPIN2016} that find no suppression with $A$.  However, as we have mentioned, the fit in Ref.~\cite{Kanazawa:2014dca} was performed before the LIR (\ref{LIRH}) was derived.  
Using both the EOMR~\eqref{e:EOM_FF} and LIR~\eqref{LIRH} we can write
\begin{align}
\frac{2} {z}\int_z^\infty\! \frac{dz_1} {z_1^2}\frac{1} {\left(\frac{1} {z}-\frac{1} {z_{1}}\right)^{\!2}}\, \hat{H}_{FU}^{\pi/c,\Im}(z,z_{1}) = H_1^{\perp(1),\pi/c}(z)+z\frac{dH_1^{\perp(1),\pi/c}(z)} {dz} - \frac{1} {z}\tilde{H}^{\pi/c}(z). 
\label{e:3rd-term}
\end{align}

\begin{figure}[ht]
\centering
\includegraphics[width=8.cm]{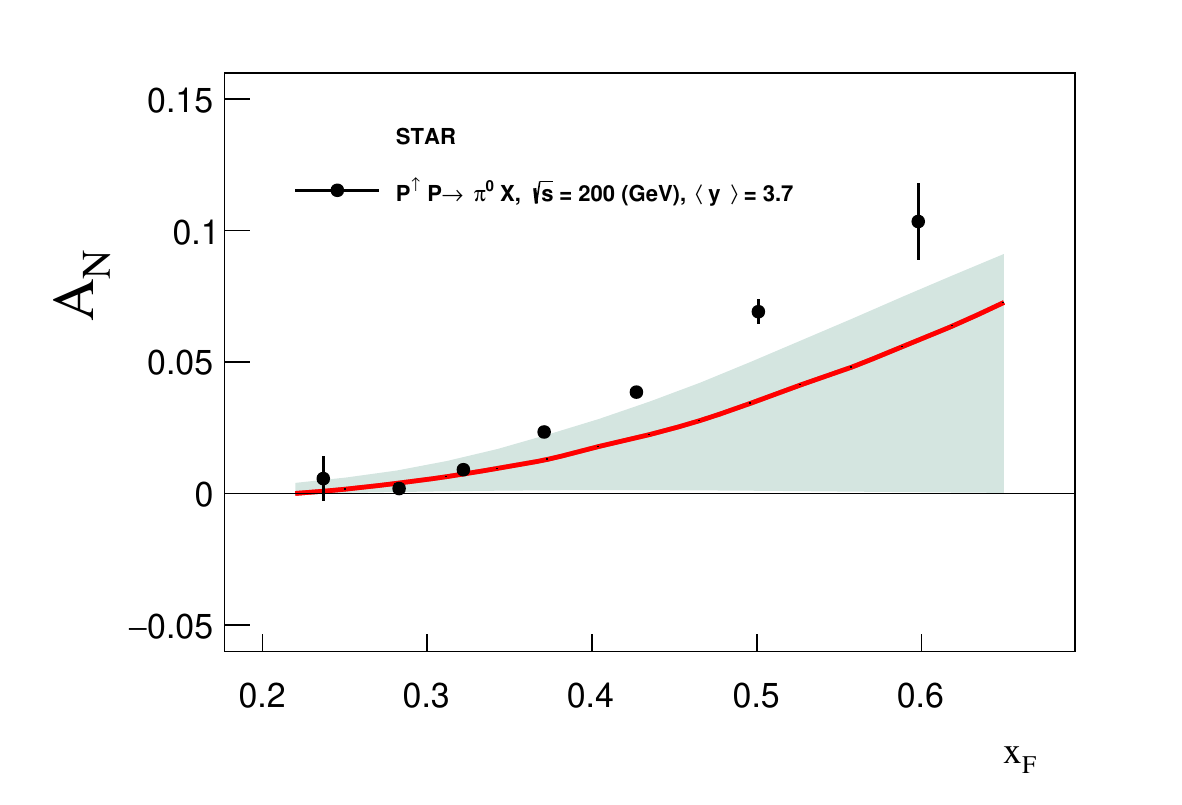} 
\caption{
STAR Collaboration data on $A_N$ for $\pi^0$  as function of $x_F$ at $\langle y \rangle = 3.7$ for positive $x_F$. The solid line and the shaded area correspond
to calculations using Eq.~\eqref{e:mathcalH3rd}. }  
\label{fig:star_pi0_third_term}
\end{figure}

\noindent With Eq.~\eqref{e:3rd-term} in hand, along with using the known input from the Collins function for $H_1^{\perp(1),q}(z)$, we can obtain a new estimate for the contribution of the third 
 term in Eq.~(\ref{e:sigmaFrag}) to $A_N$. To be specific,  we replace  $\mathcal{H}^{\pi/c,i}(x,x',z)$ in Eq.~(\ref{e:Num}) with
\begin{equation}
\mathcal{H}^{\pi/c,i}(x,x',z)\bigg|_{3^{rd}\,term\;in\;(\ref{e:sigmaFrag})} = 
\left[H_1^{\perp(1),\pi/c}(z)+z\frac{dH_1^{\perp(1),\pi/c}(z)} {dz} - \frac{1} {z}\tilde{H}^{\pi/c}(z)\right] \frac{S_{\hat{H}_{FU}}^{i}}{-x'\hat{t}-x\hat{u}}\,, \label{e:mathcalH3rd} 
\end{equation}
 where, as before, $S^i_{\hat{H}_{FU}}$ can be found in Appendix A of Ref.~\cite{Metz:2012ct}, and we include only terms involving $H_1^{\perp(1),\pi/c}(z)$.\footnote{This of course is not a complete calculation because we still must include/fit $\tilde{H}(z)$.}
 Our estimate is shown in Fig.~\ref{fig:star_pi0_third_term}. We see that the contribution to $A_N$ from the third term in Eq.~(\ref{e:sigmaFrag}) is actually moderate in size and certainly not negligible.  Since this part of $A_N$ in $p^\uparrow A$ collisions is proportional to $A^0$~\cite{Hatta:2016khv}, the fragmentation term for $A_N$ is not inconsistent with the STAR data~\cite{Heppelmann:SPIN2016} on the asymmetry in $p^\uparrow A\to \pi^0\,X$. Recently,  the author of Ref.~\cite{Zhou:2017sdx} has found that, in contrast to Ref.~\cite{Hatta:2016wjz}, the QS piece contribution to $A_N$ in $pA$ collisions is small (and may very well vanish), leaving the fragmentation term as the only source of $A_N$ in $pA$ collisions~\footnote{This conclusion is reached within the so-called hybrid approach of twist-3 and color glass condensate~\cite{Hatta:2016wjz,Zhou:2017sdx}.}.

\section{Summary \label{s:sum}} 
We have given an estimate of $A_N$ in $p^\uparrow p\to\pi\,X$ due to the kinematical twist-3 functions, which are first ($k_T$ or $p_\perp$) moments of TMD functions.  Using the newly derived LIR (\ref{LIRH}) from Ref.~\cite{Kanazawa:2015ajw}, we have written $A_N$ in terms of a maximum number of these correlators, which helps optimize the phenomenology.  In particular, we computed the  terms in $A_N$ that involve $H_1^{\perp(1)}(z)$, using the most recent TMD evolved extraction of that function~\cite{Kang:2015msa} and confirmed the conclusion of Ref.~\cite{Kanazawa:2014dca} that the fragmentation piece can be the dominant source of the asymmetry.  Note that the dynamical twist-3 FF $\hat{H}_{FU}^{\Im}(z,z_1)$, which played a key role in the analysis of Ref.~\cite{Kanazawa:2014dca}, is also crucial to our work here since $H_1^{\perp(1)}(z)$ (and $\tilde{H}(z)$) is an integral of this function.  We have simply ``re-shuffled'' the pieces by incorporating LIRs (and EOMRs) in order to write the cross section in terms of a maximal number of known functions.  We also found that the existing extractions of the QS function lead to a negligible contribution to $A_N$, although one must be mindful that this function is not well-constrained at large $x$.  Our results give a reasonable description of the data, although the central curves cannot fully account for all of the asymmetry.  However, we know the pieces that remain from the dynamical twist-3 functions, in particular $\hat{H}_{FU}^\Im(z,z_1)$ (via $\tilde{H}(z)$), must be nonzero.  Therefore, we must fit this function to $A_N$ in order to obtain a complete result for this observable.  In addition, we can reduce the rather large theoretical uncertainties by including both the transversity and Sivers functions simultaneously in such a fit.  We leave both of these matters for future work, which are needed to definitively resolve the puzzle of what causes $A_N$.  
 
\section*{Acknowledgments}
 This material is based upon work supported by the
U.S. Department of Energy, Office of Science, Office of Nuclear
Physics under Award No. DE-FG02-07ER41460 (L.G.), No.~DE-AC05-06OR23177 (A.P.), by the National Science Foundation 
under Contract No. PHY-1623454 (A.P.), and within the 
framework of the TMD Topical Collaboration (L.G., D.P., A.P.).
\bibliographystyle{h-physrev}

\begin{thebibliography}{10}

\bibitem{Pitonyak:2016hqh}
D.~Pitonyak,
\newblock Int. J. Mod. Phys. {\bf A31}, 1630049 (2016), 1608.05353.

\bibitem{Aschenauer:2016our}
E.-C. Aschenauer {\em et~al.},
\newblock (2016), 1602.03922.

\bibitem{Efremov:1981sh}
A.~V. Efremov and O.~V. Teryaev,
\newblock Sov.~J.~Nucl.~Phys. {\bf 36}, 140 (1982).

\bibitem{Efremov:1984ip}
A.~V. Efremov and O.~V. Teryaev,
\newblock Phys.~Lett. {\bf B150}, 383 (1985).

\bibitem{Qiu:1991pp}
J.-W. Qiu and G.~Sterman,
\newblock Phys.~Rev.~Lett. {\bf 67}, 2264 (1991).

\bibitem{Qiu:1991wg}
J.-W. Qiu and G.~Sterman,
\newblock Nucl.~Phys. {\bf B378}, 52 (1992).

\bibitem{Qiu:1998ia}
J.-W. Qiu and G.~Sterman,
\newblock Phys.~Rev. {\bf D59}, 014004 (1998), hep-ph/9806356.

\bibitem{Kanazawa:2000hz}
Y.~Kanazawa and Y.~Koike,
\newblock Phys.~Lett. {\bf B478}, 121 (2000), hep-ph/0001021.

\bibitem{Kouvaris:2006zy}
C.~Kouvaris, J.-W. Qiu, W.~Vogelsang, and F.~Yuan,
\newblock Phys.~Rev. {\bf D74}, 114013 (2006), hep-ph/0609238.

\bibitem{Eguchi:2006qz}
H.~Eguchi, Y.~Koike, and K.~Tanaka,
\newblock Nucl. Phys. {\bf B752}, 1 (2006), hep-ph/0604003.

\bibitem{Eguchi:2006mc}
H.~Eguchi, Y.~Koike, and K.~Tanaka,
\newblock Nucl.~Phys. {\bf B763}, 198 (2007), hep-ph/0610314.

\bibitem{Koike:2006qv}
Y.~Koike and K.~Tanaka,
\newblock Phys. Lett. {\bf B646}, 232 (2007), hep-ph/0612117.

\bibitem{Koike:2007rq}
Y.~Koike and K.~Tanaka,
\newblock Phys.Rev. {\bf D76}, 011502 (2007), hep-ph/0703169.

\bibitem{Zhou:2008fb}
J.~Zhou, F.~Yuan, and Z.-T. Liang,
\newblock Phys. Rev. {\bf D78}, 114008 (2008), 0808.3629.

\bibitem{Koike:2009ge}
Y.~Koike and T.~Tomita,
\newblock Phys.~Lett. {\bf B675}, 181 (2009), 0903.1923.

\bibitem{Yuan:2009dw}
F.~Yuan and J.~Zhou,
\newblock Phys.~Rev.~Lett. {\bf 103}, 052001 (2009), 0903.4680.

\bibitem{Kang:2010zzb}
Z.-B. Kang, F.~Yuan, and J.~Zhou,
\newblock Phys.~Lett. {\bf B691}, 243 (2010), 1002.0399.

\bibitem{Beppu:2010qn}
H.~Beppu, Y.~Koike, K.~Tanaka, and S.~Yoshida,
\newblock Phys.~Rev. {\bf D82}, 054005 (2010), 1007.2034.

\bibitem{Metz:2010xs}
A.~Metz and J.~Zhou,
\newblock Phys. Lett. {\bf B700}, 11 (2011), 1006.3097.

\bibitem{Koike:2011mb}
Y.~Koike and S.~Yoshida,
\newblock Phys.~Rev. {\bf D84}, 014026 (2011), 1104.3943.

\bibitem{Koike:2011nx}
Y.~Koike and S.~Yoshida,
\newblock Phys. Rev. {\bf D85}, 034030 (2012), 1112.1161.

\bibitem{Metz:2012ct}
A.~Metz and D.~Pitonyak,
\newblock Phys.~Lett. {\bf B723}, 365 (2013), 1212.5037.

\bibitem{Kanazawa:2013uia}
K.~Kanazawa and Y.~Koike,
\newblock Phys.~Rev. {\bf D88}, 074022 (2013), 1309.1215.

\bibitem{Beppu:2013uda}
H.~Beppu, K.~Kanazawa, Y.~Koike, and S.~Yoshida,
\newblock Phys.~Rev. {\bf D89}, 034029 (2014), 1312.6862.

\bibitem{Kanazawa:2014nea}
K.~Kanazawa, Y.~Koike, A.~Metz, and D.~Pitonyak,
\newblock Phys. Rev. {\bf D91}, 014013 (2015), 1410.3448.

\bibitem{Kanazawa:2015ajw}
K.~Kanazawa, Y.~Koike, A.~Metz, D.~Pitonyak, and M.~Schlegel,
\newblock Phys. Rev. {\bf D93}, 054024 (2016), 1512.07233.

\bibitem{Koike:2015zya}
Y.~Koike, K.~Yabe, and S.~Yoshida,
\newblock Phys. Rev. {\bf D92}, 094011 (2015), 1509.06830.

\bibitem{Klem:1976ui}
R.~D. Klem {\em et~al.},
\newblock Phys. Rev. Lett. {\bf 36}, 929 (1976).

\bibitem{Bunce:1976yb}
G.~Bunce {\em et~al.},
\newblock Phys. Rev. Lett. {\bf 36}, 1113 (1976).

\bibitem{Adams:1991rw}
E581, D.~L. Adams {\em et~al.},
\newblock Phys. Lett. {\bf B261}, 201 (1991).

\bibitem{Krueger:1998hz}
K.~Krueger {\em et~al.},
\newblock Phys. Lett. {\bf B459}, 412 (1999).

\bibitem{Allgower:2002qi}
C.~E. Allgower {\em et~al.},
\newblock Phys. Rev. {\bf D65}, 092008 (2002).

\bibitem{Adams:2003fx}
STAR, J.~Adams {\em et~al.},
\newblock Phys.~Rev.~Lett. {\bf 92}, 171801 (2004), hep-ex/0310058.

\bibitem{Adler:2005in}
PHENIX, S.~S. Adler {\em et~al.},
\newblock Phys.~Rev.~Lett. {\bf 95}, 202001 (2005), hep-ex/0507073.

\bibitem{Lee:2007zzh}
BRAHMS, J.~H. Lee and F.~Videbaek,
\newblock AIP Conf. Proc. {\bf 915}, 533 (2007),
\newblock [,533(2007)].

\bibitem{Abelev:2008af}
STAR, B.~I. Abelev {\em et~al.},
\newblock Phys. Rev. Lett. {\bf 101}, 222001 (2008), 0801.2990.

\bibitem{Arsene:2008aa}
BRAHMS, I.~Arsene {\em et~al.},
\newblock Phys. Rev. Lett. {\bf 101}, 042001 (2008), 0801.1078.

\bibitem{Adamczyk:2012qj}
STAR, L.~Adamczyk {\em et~al.},
\newblock Phys. Rev. {\bf D86}, 032006 (2012), 1205.2735.

\bibitem{Adamczyk:2012xd}
STAR Collaboration, L.~Adamczyk {\em et~al.},
\newblock Phys.~Rev. {\bf D86}, 051101 (2012), 1205.6826.

\bibitem{Bland:2013pkt}
AnDY Collaboration, L.~Bland {\em et~al.},
\newblock (2013), 1304.1454.

\bibitem{Adare:2013ekj}
PHENIX Collaboration, A.~Adare {\em et~al.},
\newblock (2013), 1312.1995.

\bibitem{Adare:2014qzo}
PHENIX Collaboration, A.~Adare {\em et~al.},
\newblock (2014), 1406.3541.

\bibitem{Airapetian:2013bim}
HERMES Collaboration, A.~Airapetian {\em et~al.},
\newblock Phys.~Lett. {\bf B728}, 183 (2014), 1310.5070.

\bibitem{Allada:2013nsw}
Jefferson Lab Hall A Collaboration, K.~Allada {\em et~al.},
\newblock Phys.~Rev. {\bf C89}, 042201 (2014), 1311.1866.

\bibitem{Boer:2003cm}
D.~Boer, P.~J. Mulders, and F.~Pijlman,
\newblock Nucl.~Phys. {\bf B667}, 201 (2003), hep-ph/0303034.

\bibitem{Sivers:1989cc}
D.~W. Sivers,
\newblock Phys.~Rev. {\bf D41}, 83 (1990).

\bibitem{Kang:2011hk}
Z.-B. Kang, J.-W. Qiu, W.~Vogelsang, and F.~Yuan,
\newblock Phys.~Rev. {\bf D83}, 094001 (2011), 1103.1591.

\bibitem{Kang:2012xf}
Z.-B. Kang and A.~Prokudin,
\newblock Phys.Rev. {\bf D85}, 074008 (2012), 1201.5427.

\bibitem{Metz:2012ui}
A.~Metz {\em et~al.},
\newblock Phys.~Rev. {\bf D86}, 094039 (2012), 1209.3138.

\bibitem{Kanazawa:2010au}
K.~Kanazawa and Y.~Koike,
\newblock Phys.~Rev. {\bf D82}, 034009 (2010), 1005.1468.

\bibitem{Kanazawa:2011bg}
K.~Kanazawa and Y.~Koike,
\newblock Phys.~Rev. {\bf D83}, 114024 (2011), 1104.0117.

\bibitem{Zhou:2008mz}
J.~Zhou, F.~Yuan, and Z.-T. Liang,
\newblock Phys.~Rev. {\bf D79}, 114022 (2009), 0812.4484.

\bibitem{Kang:2008ey}
Z.-B. Kang and J.-W. Qiu,
\newblock Phys.~Rev. {\bf D79}, 016003 (2009), 0811.3101.

\bibitem{Kang:2012em}
Z.-B. Kang and J.-W. Qiu,
\newblock Phys.~Lett. {\bf B713}, 273 (2012), 1205.1019.

\bibitem{Boer:1997nt}
D.~Boer and P.~J. Mulders,
\newblock Phys. Rev. {\bf D57}, 5780 (1998), hep-ph/9711485.

\bibitem{Kanazawa:2000kp}
Y.~Kanazawa and Y.~Koike,
\newblock Phys. Lett. {\bf B490}, 99 (2000), hep-ph/0007272.

\bibitem{Collins:1992kk}
J.~C. Collins,
\newblock Nucl.~Phys. {\bf B396}, 161 (1993), hep-ph/9208213.

\bibitem{Anselmino:2012rq}
M.~Anselmino {\em et~al.},
\newblock Phys.~Rev. {\bf D86}, 074032 (2012), 1207.6529.

\bibitem{Kanazawa:2014dca}
K.~Kanazawa, Y.~Koike, A.~Metz, and D.~Pitonyak,
\newblock Phys. Rev. {\bf D89}, 111501(R) (2014), 1404.1033.

\bibitem{Anselmino:2008sga}
M.~Anselmino {\em et~al.},
\newblock Eur.~Phys.~J. {\bf A39}, 89 (2009), 0805.2677.

\bibitem{Anselmino:2013vqa}
M.~Anselmino {\em et~al.},
\newblock Phys.~Rev. {\bf D87}, 094019 (2013), 1303.3822.

\bibitem{Owens:1986mp}
J.~F. Owens,
\newblock Rev. Mod. Phys. {\bf 59}, 465 (1987).

\bibitem{Kang:2013ufa}
Z.-B. Kang, I.~Vitev, and H.~Xing,
\newblock Phys. Rev. {\bf D88}, 054010 (2013), 1307.3557.

\bibitem{Brodsky:2002cx}
S.~J. Brodsky, D.~S. Hwang, and I.~Schmidt,
\newblock Phys. Lett. {\bf B530}, 99 (2002), hep-ph/0201296.

\bibitem{Collins:2002kn}
J.~C. Collins,
\newblock Phys.~Lett. {\bf B536}, 43 (2002), hep-ph/0204004.

\bibitem{Politzer:1980me}
H.~D. Politzer,
\newblock Nucl. Phys. {\bf B172}, 349 (1980).

\bibitem{Boer:1997bw}
D.~Boer, P.~J. Mulders, and O.~V. Teryaev,
\newblock Phys. Rev. {\bf D57}, 3057 (1998), hep-ph/9710223.

\bibitem{Bacchetta:2006tn} 
  A.~Bacchetta, M.~Diehl, K.~Goeke, A.~Metz, P.~J.~Mulders and M.~Schlegel,
 \newblock JHEP {\bf 0702}, 093 (2007),
  hep-ph/0611265.

\bibitem{Boer:1997mf}
D.~Boer, R.~Jakob, and P.~J. Mulders,
\newblock Nucl. Phys. {\bf B504}, 345 (1997), hep-ph/9702281.

\bibitem{Lu:2015wja}
Z.~Lu and I.~Schmidt,
\newblock Phys. Lett. {\bf B747}, 357 (2015), 1501.04379.

\bibitem{Wang:2016tix}
X.~Wang and Z.~Lu,
\newblock Phys. Rev. {\bf D93}, 074009 (2016), 1601.01574.

\bibitem{Kodaira:1998jn}
J.~Kodaira and K.~Tanaka,
\newblock Prog. Theor. Phys. {\bf 101}, 191 (1999), hep-ph/9812449.

\bibitem{Belitsky:1997ay}
A.~V. Belitsky,
\newblock {Leading order analysis of the twist-three space-like and time-like
  cut vertices in QCD},
\newblock in {\em {Annual 31st PNPI Winter School on Nuclear and Particle
  Physics St. Petersburg, Russia, February 24-March 2, 1997}}, 1997,
  hep-ph/9703432.

\bibitem{Accardi:2009au}
A.~Accardi, A.~Bacchetta, W.~Melnitchouk, and M.~Schlegel,
\newblock JHEP {\bf 11}, 093 (2009), 0907.2942.

\bibitem{Kang:2012ns}
Z.-B. Kang, I.~Vitev, and H.~Xing,
\newblock Phys. Rev. {\bf D87}, 034024 (2013), 1212.1221.

\bibitem{Anselmino:2007fs}
M.~Anselmino {\em et~al.},
\newblock Phys.~Rev. {\bf D75}, 054032 (2007), hep-ph/0701006.

\bibitem{Anselmino:2008jk}
M.~Anselmino {\em et~al.},
\newblock Nucl. Phys. Proc. Suppl. {\bf 191}, 98 (2009), 0812.4366.

\bibitem{Anselmino:2012aa}
M.~Anselmino, M.~Boglione, and S.~Melis,
\newblock (2012), 1204.1239.

\bibitem{Anselmino:2013rya}
M.~Anselmino {\em et~al.},
\newblock Phys.~Rev. {\bf D88}, 054023 (2013), 1304.7691.

\bibitem{Echevarria:2014xaa}
M.~G. Echevarria, A.~Idilbi, Z.-B. Kang, and I.~Vitev,
\newblock Phys.~Rev. {\bf D89}, 074013 (2014), 1401.5078.

\bibitem{Anselmino:2015sxa}
M.~Anselmino {\em et~al.},
\newblock Phys. Rev. {\bf D92}, 114023 (2015), 1510.05389.

\bibitem{Kang:2015msa}
Z.-B. Kang, A.~Prokudin, P.~Sun, and F.~Yuan,
\newblock Phys. Rev. {\bf D93}, 014009 (2016), 1505.05589.

\bibitem{Wandzura:1977qf}
S.~Wandzura and F.~Wilczek,
\newblock Phys. Lett. {\bf B72}, 195 (1977).

\bibitem{Metz:2008ib}
A.~Metz, P.~Schweitzer, and T.~Teckentrup,
\newblock Phys. Lett. {\bf B680}, 141 (2009), 0810.5212.

\bibitem{Teckentrup:2009tk}
T.~Teckentrup, A.~Metz, and P.~Schweitzer,
\newblock Mod. Phys. Lett. {\bf A24}, 2950 (2009), 0910.2567.

\bibitem{Collins:2011zzd}
J.~Collins,
\newblock {\em Foundations of Perturbative QCD}, (Cambridge University Press,
  Cambridge, England, 2011).

\bibitem{Aybat:2011ge}
S.~M. Aybat, J.~C. Collins, J.-W. Qiu, and T.~C. Rogers,
\newblock Phys.~Rev. {\bf D85}, 034043 (2012), 1110.6428.

\bibitem{Kang:2011mr}
Z.-B. Kang, B.-W. Xiao, and F.~Yuan,
\newblock Phys.~Rev.~Lett. {\bf 107}, 152002 (2011), 1106.0266.

\bibitem{Sun:2013hua}
P.~Sun and F.~Yuan,
\newblock Phys.~Rev. {\bf D88}, 114012 (2013), 1308.5003.

\bibitem{Kang:2010xv}
Z.-B. Kang,
\newblock Phys.~Rev. {\bf D83}, 036006 (2011), 1012.3419.

\bibitem{Gamberg:2013kla}
L.~Gamberg, Z.-B. Kang, and A.~Prokudin,
\newblock Phys.Rev.Lett. {\bf 110}, 232301 (2013), 1302.3218.

\bibitem{Dudek:2012vr}
J.~Dudek {\em et~al.},
\newblock Eur.~Phys.~J. {\bf A48}, 187 (2012), 1208.1244.

\bibitem{Ye:2016prn}
Z.~Ye {\em et~al.},
\newblock (2016), 1609.02449.

\bibitem{Heppelmann:2013ewa}
STAR, S.~Heppelmann,
\newblock PoS {\bf DIS2013}, 240 (2013).

\bibitem{Hatta:2016khv}
Y.~Hatta, B.-W. Xiao, S.~Yoshida, and F.~Yuan,
\newblock Phys. Rev. {\bf D95}, 014008 (2017), 1611.04746.

\bibitem{Kang:2011ni}
Z.-B. Kang and F.~Yuan,
\newblock Phys. Rev. {\bf D84}, 034019 (2011), 1106.1375.

\bibitem{Heppelmann:SPIN2016}
STAR, S.~Heppelmann,
\newblock Talk at SPIN 2016.

\bibitem{Zhou:2017sdx} 
  J.~Zhou,
\newblock  1704.04901.

\bibitem{Hatta:2016wjz} 
  Y.~Hatta, B.~W.~Xiao, S.~Yoshida and F.~Yuan,
\newblock  Phys.\ Rev.\ {\bf D94}, 054013 (2016), 1606.08640.
  
\end{thebibliography}

\end{document}